\documentclass[prd,aps,amsfonts,showpacs,superscriptaddress,preprintnumbers,nofootinbib,10pt]{revtex4-2}

\usepackage{tikz}
\usetikzlibrary{shapes,arrows,positioning}
\usepackage{graphicx}
\usepackage{xcolor}
\usepackage{rotating}
\usepackage{bm,amsmath,amssymb}
\usepackage[utf8]{inputenc}
\usepackage{footmisc}
\usepackage{diagbox}
\usepackage{soul}
\usepackage[colorlinks=true, urlcolor=blue, linkcolor=blue, citecolor=blue, hyperindex=true, linktocpage=true]{hyperref}

\newcommand{\arcsinh}{\text{arcsinh}}
\renewcommand{\epsilon}{\varepsilon}
\newcommand{\const}{{\text{const}}}

\newcommand{\ideal}{{\text{ideal}}}

\newcommand{\Ckappasigma}{{C_{\kappa_1}}}
\newcommand{\Ctaupi}{{C_{\tau_\pi}}}
\newcommand{\mzero}{{m_0}}

\newcommand{\Czeta}{{C_\zeta}}

\newcommand{\mn}{{\mu\nu}}

\newcommand{\cond}{{\sigma}}
\newcommand{\mass}{{m_0}}

\begin{document}

\title{Superfluids in expanding backgrounds and attractor times}

%\title{Superfluids in expanding backgrounds and attractor times}

\author{Guri K.~Buza}
\email[]{guri.buza@fmf.uni-lj.si}
 \affiliation{Faculty of Mathematics and Physics, University of Ljubljana, Jadranska ulica 19, SI-1000 Ljubljana, Slovenia}

\author{Toshali Mitra}
\email[]{toshali.mitra@apctp.org}
\affiliation{Asia Pacific Center for Theoretical Physics, Pohang, 37673, Korea}
\affiliation{Department of Physics, Pohang University of Science and Technology, Pohang, 37673, Korea}

\author{Alexander Soloviev}
\email[]{alexander.soloviev@fmf.uni-lj.si}
\affiliation{Faculty of Mathematics and Physics, University of Ljubljana, Jadranska ulica 19, SI-1000 Ljubljana, Slovenia}

\begin{abstract}
We determine the behavior of an out-of-equilibrium superfluid, composed of a $U(1)$ Goldstone mode coupled to hydrodynamic modes in a M\" uller-Israel-Stewart theory, in expanding backgrounds relevant to heavy ion collision experiments and cosmology. For suitable initial conditions, the evolution of the hydrodynamic variables leads to a change in the potential of the Goldstone mode, spontaneously breaking the symmetry. After some time, the condensate becomes small, leading the system evolution to be well described via hydrodynamic attractors for a timescale that we determine
in Bjorken and Gubser flows. We define this new timescale as the \textit{attractor time} and show its dependence on initial conditions. In the case of the Gubser flow, we provide for the first time a complete description of the nonlinear evolution of the system, including a novel nonlinear regime of constant anisotropy not found in the Bjorken evolution.
Finally, we consider the superfluid in the dynamical FLRW (Friedmann-Lemaitre-Roberston-Walker) background, where we observe a similar attractor behavior, dependent on the initial conditions, that at late times approaches a regime dominated by the condensate.
\end{abstract}

\maketitle

\section{Introduction}

Superfluidity is a ubiquitous phenomenon found in diverse fields of physics, including high energy particle physics~\cite{Alford:2007xm,Schafer:2009dj}, condensed matter systems such as cold atoms~\cite{kapitza1938viscosity} and the description of astrophysical objects such as neutron stars~\cite{Haskell:2017lkl}. Recently, there has been considerable interest in promoting the Goldstone mode to a state parameter to study the interplay between such modes and hydrodynamic modes, for example in the case of the chiral phase transition \cite{Grossi:2020ezz,Grossi:2021gqi,Florio:2021jlx,Florio:2023kmy} as it may prove relevant in the search for the QCD critical point at the Beam Energy Scan experiment~\cite{Du:2024wjm}. Moreover, in certain systems, such as in heavy ion  collisions with approximately boost-invariant flows \cite{Romatschke:2007mq}, small systems of strongly interacting fermions \cite{Brandstetter:2023jsy} or time dependent scattering length in cold atoms \cite{Fujii:2024yce}, it has been observed that although the system is far from equilibrium, hydrodynamics remains a remarkably good description of the system outside its naive range of validity, which can be explained in part due to the presence of hydrodynamic attractors (for reviews, see
\cite{Soloviev:2021lhs,Jankowski:2023fdz}). Thus, the question naturally arises of how hydrodynamic attractors change in the vicinity of a superfluid phase transition. 

The hydrodynamic degrees of freedom of a normal fluid are the usual conserved densities such as temperature, fluid velocity, and charge densities. However, in the presence of broken symmetry, such as spontaneously broken $U(1)$ symmetry, the order parameter enters the dynamics of the system. These new degrees of freedom are those of the superfluid phases, where superfluidity can be defined as a phase of a system with spontaneously broken continuous symmetry below a certain critical temperature \cite{Son:2000ht, Son:2002zn,schmitt2014IntroductionSuperfluidityFieldtheoretical}.  

\begin{figure}[ht]
    \centering
    \includegraphics[width=0.5\textwidth]{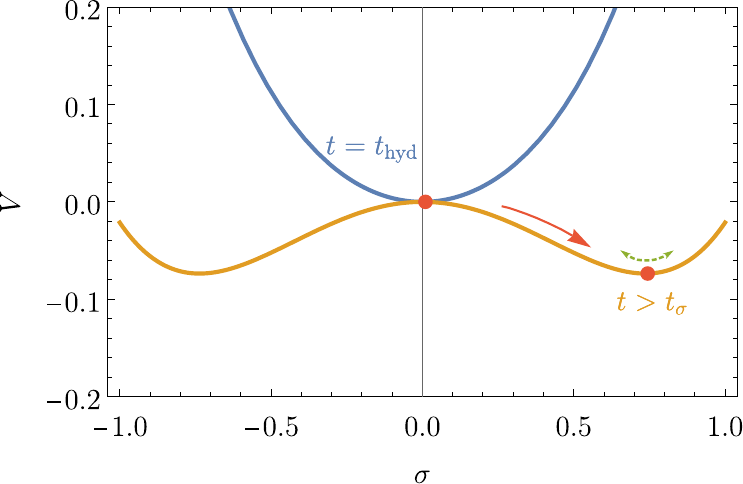}
    \caption{The condensate (red point) in the potential \eqref{potential} at early times (blue) in the unbroken phase and for later times (orange) in the broken phase. Due to the dependence of the mass term on the dynamical energy density, the system undergoes a phase transition. At later times, the condensate can undergo oscillation as well (green) before it settles at the bottom of the potential well.}
    \label{fig:potential_visualization}
\end{figure}

In this work, we employ the formalism of \cite{Mitra:2020hbj}, which builds on the Son-Nicolis approach \cite{Son:2002zn,Nicolis:2011cs}. The central premise is that we dynamically couple a $U(1)$ scalar field to a fluid in an expanding background, whose dissipation is governed by the M\" uller-Israel-Stewart (MIS)  framework \cite{mueller,israel}. The potential of the scalar field is chosen such that the mass term is dependent on the energy density of the fluid, changing sign as the system passes a critical value. 
The expansion of the background metric cools the system, eventually leading to the symmetry of the scalar field to be spontaneously broken, resulting in a transition from the normal fluid phase to the superfluid phase (see Fig~\ref{fig:potential_visualization}).

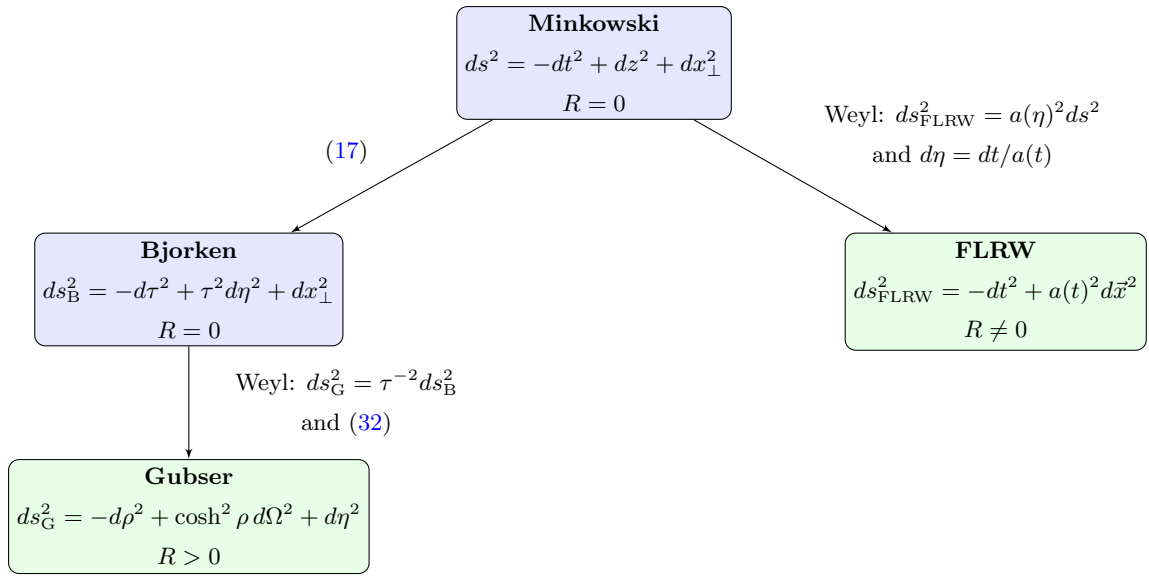
\begin{figure}
\begin{tikzpicture}[auto, node distance=2cm, >=latex']
    % Define styles
    \tikzstyle{flat}=[rectangle, draw, rounded corners, align=center, fill=blue!10]
    \tikzstyle{curved}=[rectangle, draw, rounded corners, align=center, fill=green!10]
    \tikzstyle{arrow} = [->,>=latex']
    \tikzstyle{weyl} = [->,>=latex', solid, color=black]

    % Nodes
    \node[flat] (minkowski) {\textbf{Minkowski} \\[0.5em] $ds^2 = -dt^2 + dz^2 + dx_\perp^2$ \\[0.5em] $R = 0$}; 
    
    \node[flat, below left=1.5cm and 1.5cm of minkowski] (bjorken) 
    {\textbf{Bjorken} \\[0.5em] $ds^2_{\rm B} = -d\tau^2 + \tau^2 d\eta^2 + dx_\perp^2$ \\[0.5em] $R = 0$};
    
    \node[curved, below=1.5cm of bjorken] (gubser) 
    {\textbf{Gubser} \\[0.5em] $ds^2_{\rm G} = -d\rho^2 + \cosh^2 \rho\, d\Omega^2 + d\eta^2$ \\[0.5em] $R > 0$};
    
    \node[curved, below right=1.5cm and 1.5cm of minkowski] (flrw) 
    {\textbf{FLRW} \\[0.5em] $ds^2_{\rm FLRW} = -dt^2 + a(t)^2 d\vec{x}^2$ \\[0.5em] $R \neq 0$};
    
    % Arrows with labels
    \draw[arrow] (minkowski) -- node[align=center, xshift=-1cm, yshift=0.6cm] {\small \eqref{bjork-transf}} (bjorken);
    
    \draw[weyl] (bjorken) -- node[align=center, xshift=0.5cm, yshift=0cm] {Weyl: $ds^2_{\rm G} = \tau^{-2} ds^2_{\rm B}$ \\[0.5em] and \eqref{gubser-transf}} (gubser);
    
    \draw[weyl] (minkowski) -- node[align=center, xshift=0.3cm] {Weyl: $ds^2_{\rm FLRW} = a(\eta)^2 ds^2$ \\[0.5em] and $d\eta = dt / a(t)$} (flrw);
\end{tikzpicture}
\caption{Overview showing the relationships between the different backgrounds explored here. $R$ denotes the Ricci scalar and serves as an indication for the curvature of the spacetime.}\label{fig:metrics}
\end{figure}

More concretely, we study the boost-invariant Bjorken \cite{Bjorken:1982qr} and Gubser flow \cite{Gubser:2010ze,Gubser:2010ui} (see also the related exact solutions in boost-invariant superfluid flows \cite{Rodgers:2022fuv}), both of which have seen successful phenomenological exploitation in heavy ion physics \cite{Romatschke:2017ejr,Giacalone:2019ldn}, as well as the  Friedmann–Lema\^ itre–Robertson–Walker (FLRW) background \cite{Wald:1984rg}, a cornerstone for describing the evolution of the early universe \cite{Hinshaw_2013}. For a guide to the metrics used, see Fig.~\ref{fig:metrics}. In these backgrounds, depending on the initial conditions, we see that the evolution of the superfluid is dictated by the hydrodynamic attractor with unbroken symmetry at intermediate times, while at late times the system approaches one of the symmetry breaking fixed points.

A key result of this work is the novel notion of attractor time, the interval of time that the dynamics of a system is governed primarily by the attractor solution. Concretely, for the Bjorken flow, this occurs when the system is initially in the unbroken phase (for systems with sufficiently large initial temperatures compared to the critical temperature, $T_0>T_c$) and the condensate is at the minimum of the potential, $\cond=0$. As the condensate dynamics essentially decouple, the system is governed by the hydrodynamic equations in an expanding background, which is denoted as the hydrodynamic attractor. Eventually, as the system cools and the temperature drops, the potential transits to the broken superfluid phase and the condensate rapidly drops into a minimum. Thereafter, the evolution is no longer governed by the attractor, but rather by the fixed points with constant temperature and condensate. The existence of these fixed points is possible because the non-zero condensate lowers the energy with respect to the vacuum, allowing the expanding system to maintain a self-consistent constant temperature.

The usefulness of such a timescale can be demonstrated by considering a typical flow used to describe collider experiments. In the typical example of a (normal) Bjorken fluid, the time the system is well described by the hydrodynamic attractor is infinite. However, the physical system fails to have a hydrodynamic description as it freezes out and undergoes hadronization at some finite time, indicating that the system has deviated from the hydrodynamic attractor. Thus, our present model of expanding superfluid flow
 provides a picture of a fluid transitioning to a non-hydrodynamic description.

Another important set of results is the first description of the evolution of the superfluid Gubser flow. The evolution is qualitatively distinct to Bjorken flow, due in part to the difference in the expansion rate as a function of system time:
\begin{align}
    \nabla\cdot u = \begin{cases}
        1/\tau, & \text{Bjorken flow},\\
        2\tanh \rho, & \text{Gubser flow},
    \end{cases}
\end{align}
where $u^\mu$ is the fluid velocity, $\tau =\sqrt{t^2-z^2}$ is the proper time in Milne coordinates and $\rho$ is the Gubser time coordinate (whose relationship to Bjorken coordinates is given in \eqref{gubser-transf}). A characteristic feature of Bjorken flow is that for asymptotically large times, the expansion rate becomes small. This means that at late times, a gradient expansion is sensible. However, Gubser flow at large times has a constant expansion rate, with only a gradient expansion sensible near $\rho=0$ \cite{Dash:2020zqx}, which in Bjorken coordinates corresponds to intermediate proper time. Another key difference is that while Bjorken flow is a comoving frame which requires a transformation from Minkowski coordinates, Gubser flow requires an additional conformal Weyl transformation. In other words, Bjorken space is Ricci flat, while Gubser has positive curvature. We note in passing that Gubser flow has been predicted to be important for $pp$ systems, especially for two and four particle cumulants \cite{Taghavi:2019mqz}, with recent experimental evidence from the ALICE collaboration \cite{ALICE:2023lyr}.

The qualitative picture of such a superfluid Gubser flow (see e.g. Fig.~\ref{Fig:gubser_chi}) begins with a region initially dominated by an inviscid fluid with the condensate quickly dropping to the bottom of the potential (\textbf{Region II}). As the system expands with increasing $\rho,$ the dissipative contribution to the fluid becomes more pronounced (\textbf{Region III}) and the anisotropy approaches a fixed ratio of dimensionless constants of the shear viscosity to the relaxation time, $\chi \rightarrow \sqrt{C_\eta/C_{\tau_\pi}}.$ This is in line with the standard MIS Gubser flow picture \cite{Denicol:2018pak,Dash:2020zqx}. At some later point, the temperature drops low enough for the condensate to begin rolling down the potential well extremely slowly (\textbf{Region IV}). Similar to the previous viscous hydro regime, this part of the evolution is characterized by a constant value of the anisotropy, namely $\chi \rightarrow C_\eta/C_{\tau_\pi}.$ Finally, the condensate rapidly evolves to one of the minima of the potential at late times (\textbf{Region V}).

In both the Bjorken and Gubser flow, the background metric was not a dynamical variable in the system. Moreover, both systems exhibit features relevant to early universe cosmology, namely the early time smoothening out of inhomogeneities via the approach to the attractor \cite{Romatschke:2017acs} and late time inflation due to the exponential growth of the condensate \cite{gorbunov}. In this vein, we examined the superfluid in the FLRW background. Since gravity is dynamical in this case, we have the Hubble parameter, $H(t)$, as an additional dynamical degree of freedom. The expansion rate in the FLRW background is
\begin{align}
    \nabla\cdot u = 3 H(t).
\end{align}
However, unlike in the Bjorken or in the Gubser cases, the existence of a long-lived attractor regime depends significantly on the initial conditions, with the initial value of the condensate needing to be sufficiently small to see the attractor. {Note, here we account for backreaction between the fluid variables and the metric, which differentiates our hydrodynamic attractor analysis in FLRW from the analysis in \cite{Du_2021}.} 
An important feature we see is that as the condensate falls into the bottom of the potential well, the evolution is not complete: since both the background metric and the potential are dynamical, the scalar field oscillates around its minimum for some period of time with exponentially decreasing amplitude (see Fig.~\ref{fig:potential_visualization}). We should point out that this behavior is also seen in the Bjorken flow for low enough friction in the scalar sector, which is however exponentially suppressed for the Gubser superfluid.

Scalar fields coupled to Einstein's equation have been extensively studied in scalar field cosmology \cite{chervonea2019scalarFieldCosmology,martin2024encyclopaediainflationaris}. This work is most closely related to studies involving symmetry-breaking potentials. However, a key distinction here is that we consider the fully backreacted metric: the potential depends on the temperature. Its dynamical shape causes the scalar field to visit different symmetry-breaking regions which may or may not satisfy the slow-roll conditions at  different times during the evolution. A detailed slow-roll analysis is left for future work.

The organization of the paper is as follows. We discuss the general covariant set-up in Section~\ref{sec:setup}, recapping the discussion in \cite{Mitra:2020hbj}. We then turn our attention to the Bjorken and Gubser flow in Sections \ref{sec:bjork} and \ref{sec:gubs}, respectively. Finally, in Section~\ref{sec:tldr} we discuss a model of our universe via a dissipative superfluid in the FLRW metric. 

\section{Set up}\label{sec:setup}

We summarize our set up here. Note that it was first outlined in \cite{Mitra:2020hbj}, complete with a derivation with a more general kinetic term than we consider here and for a more general equation of state. The essential idea is that we have long wavelength degrees of freedom, described via fluid degrees of freedom, which are coupled to additional non-hydrodynamic degrees of freedom, namely those of a $U(1)$ scalar field $\Sigma$. Moreover, the scalar field undergoes symmetry breaking when the temperature falls below the critical value. The appropriate effective action is 
\begin{equation}\label{Eq:EffectiveAction}
S=\int d^4 x \sqrt{-g}\, P(T,\mu,\Sigma)=\int d^4 x \sqrt{-g}\left[ p(T,\mu) -\frac{1}{2} (D_\mu \Sigma)(D^\mu \Sigma)^\dagger - V( \Sigma, T)\right]\,,
\end{equation}
where $P(T,\mu,\Sigma)$ is the generalized pressure, $p(T,\mu)$ is the fluid pressure and the gauge covariant derivative is $D_\mu=\nabla_\mu+igA_\mu$. The complex scalar field is further decomposed into the condensate $\cond$ and the phase $\psi$, $\Sigma = \sigma e^{i\psi}$. $V$ is the symmetry breaking potential, given by
\begin{align}\label{potential}
V( \cond,  T) = \frac{m_0 (T-T_c)}{2}  \cond^2 + \frac{\lambda}{4}  \cond^4,
\end{align}
in which the critical temperature $T_c>0$, the quartic coupling $\lambda >0$, and $m_0>0$ are constants.\footnote{Note that in Ref.~\cite{Mitra:2020hbj} the scalar field is denoted by $\rho$. In this work, we denote the condensate of the scalar field by $\sigma$, and we keep the standard notation $\rho$ for the de Sitter time in the Gubser flow in Sec.~\ref{sec:gubs}.}

By varying the effective action  \cite{brown1993ActionFunctionalsRelativistic}, we can immediately determine the ideal energy-momentum tensor
\begin{align}
T^{\mu\nu}_{\rm ideal}&\equiv\frac{2}{\sqrt{-g}}\frac{\delta (\sqrt{-g}P)}{\delta g_{\mu\nu}} 
= \varepsilon\, u^\mu u^\nu + P \, \Delta^{\mu\nu}
-\left( \nabla^\mu  \cond \nabla^\nu  \cond +  \cond^2 D^\mu \psi D^\nu \psi\right)
%\varepsilon \, u^\mu u^\nu + (p-V) \, \Delta^{\mu\nu}-\left( \nabla^\mu  \cond \nabla^\nu  \cond +  \cond^2 D^\mu \psi D^\nu \psi\right)
\end{align}
where the total energy density is $\varepsilon=(T\frac{\partial}{\partial T}+\mu \frac{\partial}{\partial \mu} -1)P$, $u^\mu$ is the normalized ($u^\mu u_\mu=-1$) four velocity and the spatial projector is $\Delta^{\mu\nu} = g^{\mu\nu} + u^\mu u^\nu$. The conserved $U(1)$ current is found by varying the action with respect to the gauge field
\begin{align}
j^\mu_{\rm ideal} &\equiv\frac{1}{\sqrt{-g}}\frac{\delta( \sqrt{-g}P)}{\delta A_{\mu}} = \mathcal{N} u^\mu+  \cond^2 D^\mu \psi, 
\end{align}
 where $\mathcal{N}= \frac{\partial P}{\partial \mu}$ denotes the number density and $A_\mu u^\mu=\mu$. The entropy density at fixed chemical potential is defined by $\mathcal{S}=\frac{\partial P}{\partial T}$. Thus, we see that the standard thermodynamic identity $\varepsilon+P=T\mathcal{S}+\mu \mathcal{N}$ holds. Note that both the energy-momentum tensor and the current can be viewed as the sum of normal and coherent superfluid components \cite{alford.mallavarapu.ea2013ComplexScalarField,schmitt2014IntroductionSuperfluidityFieldtheoretical}. Following \cite{Mitra:2020hbj}, the normal components are
\begin{align}
    T^\mn_{\rm n}= \varepsilon\, u^\mu u^\nu + P \, \Delta^{\mu\nu}, \quad \quad j^\mu_{\rm n}  = \mathcal{N} u^\mu,
\end{align} while the other terms, directly proportional to the $U(1)$ scalar field (i.e.~containing $\sigma$ and $\psi$), are the superfluid components.

We are interested in studying the dissipative equations of motion. At present, we include dissipation via 
\begin{align}\label{emt}  T^{\mu\nu}&=T^{\mu\nu}_{\rm ideal}+ \Pi^{\mu\nu},
\end{align}
where $\Pi^\mn$ is the dissipative tensor. It is customary \cite{Romatschke:2017ejr} to split $\Pi^\mn$ into a transverse traceless piece, $\pi^\mn$, i.e.~$\pi^\mn g_\mn=0$ and $\pi^\mn u_\mu=0,$ and a  bulk term with nonvanishing trace 
\begin{equation}
    \Pi^\mn = \pi^\mn + \Pi \, \Delta^\mn,
\end{equation}
where the first term denotes the shear dissipation tensor, and $\Pi$ denotes the bulk pressure. The stress tensor is conserved 
\begin{align} \label{emt-conservation}
    \nabla_\mu T^\mn=0.
\end{align}

{Similarly, the ideal equations of motion for the scalar field, found by varying the action w.r.t. $\Sigma=\sigma e^{i\psi}$}, can be modified by adding dissipative sources $\theta_1$ and $\theta_2$ to the equations of motion of $ \cond$ and $\psi$ respectively, so that
\begin{align}\label{Eq:EomSigma}
\nabla_\mu \nabla^\mu  \cond  -\frac{\partial V}{\partial  \cond} -  \cond  D_\mu\psi D^\mu\psi &= \theta_1,  \\
\nabla_\mu j^\mu_\psi &= \theta_2. \label{Eq:EomSigma2}
\end{align}
As detailed in \cite{Mitra:2020hbj}, we require that the requisite constitutive relations lead to the positive definite divergence of the entropy current $\nabla_\mu (\mathcal{S} u^\mu) \geq 0$. This requirement of positive entropy production constrains the dissipative sources to be \cite{Mitra:2020hbj}
\begin{eqnarray}\label{Eq:Constitutive}
\theta_1 = -\kappa_1 (u\cdot\nabla)  \cond, &\quad \theta_2 = - \kappa_2 (u\cdot D) \psi,
\end{eqnarray}
where $\kappa_1$ and $\kappa_2$ are positive definite functions of temperature $T$ and $\mu$. 

In practice, the phase plays little role in the dynamics, quickly settling to a constant value. Thus, to simplify the presentation, throughout this work we will consider the case of zero chemical potential, $\mu=0,$ which due to the Josephson constraint sets the phase to a constant\footnote{Note that in equilibrium, entropy production is absent and the Josephson condition is satisfied $(u\cdot D) \psi = 0$. %, i.~e. $(u\cdot \nabla) \psi = \mu$.
}. We will return to this assumption in future work, particularly studying the $O(4)$ phase transition which has a more nuanced non-Abelian structure. Moreover, for the conformal Bjorken and Gubser flow in Secs. \ref{sec:bjork} and \ref{sec:gubs},  we will set $p =\varepsilon/3= T^4$. In the case of the non-conformal FLRW in Sec.~\ref{sec:tldr}, the fluid pressure will be related to the energy density via $p=w \varepsilon.$

Finally, in order to develop a MIS-type formulation, we simply replace the constitutive relations for $\pi^{\mu\nu}$ and $\Pi$
%and $j^\mu$ 
by the dynamical equations \cite{mueller,israel}
\begin{align}
\tau_\pi \left(u\cdot\nabla + \frac{4}{3} \, \nabla \cdot u\right) \pi^{\mu\nu}+ \pi^{\mu\nu} &= -  2 \, \eta \, \sigma^{\mu\nu},\label{mis-visc} \\
\tau_\Pi \, (u \cdot \nabla) \;  \Pi + \Pi &= - \, \zeta \,  \nabla \cdot u,\label{mis-visc-bulk}
%\\
%\label{Eq:MIS}
%(u\cdot\nabla)(q^\mu+ j^\mu_\psi)+ \frac{1}{\tau_q}(q^\mu+ j^\mu_\psi)
%&=  -  \frac{1}{\tau_q}\kappa \nabla^\mu \left( \frac{\mu}{T}\right),
\end{align}
where $\eta$ is the shear viscosity,  the shear tensor is $\sigma^\mn :=\frac{1}{2} \Delta^{\mu\alpha}\Delta^{\nu\beta}(\nabla_\alpha u_\beta+\nabla_\beta u_\alpha)-\frac{1}{3}\Delta^\mn\nabla_\alpha u^\alpha$ and $\zeta$ is the bulk viscosity. Since we consider the case of zero chemical potential, we omit the MIS equation for the current $j^\mu$. The shear and bulk relaxation times are given by $\tau_\pi$ and $\tau_\Pi$, respectively.
Note that we have included the BRSSS improvement term \cite{Baier:2007ix} in the evolution of $\pi^\mn$ to ensure positive entropy production. 

Thus, the key equations of motion that we will consider are the energy momentum tensor conservation \eqref{emt-conservation}, the equation of motion of the condensate \eqref{Eq:EomSigma}, and the MIS equations \eqref{mis-visc} and \eqref{mis-visc-bulk}.  Additionally, in Sec.~\ref{sec:tldr}, the metric will be dynamical, which will lead to the inclusion of Einstein's equations, which we will discuss there.

\section{Bjorken flow}\label{sec:bjork}

In this section, we revisit the superfluid in the Bjorken flow, discussed in \cite{Mitra:2020hbj}. The metric is given by
\begin{align}
    ds^2_{\rm B}=-d\tau^2+dx_\perp^2+\tau^2 d\eta^2,
\end{align}
where $x_\perp$ denotes the transverse directions and $\eta$ is the rapidity. These are related to the Minkowski space coordinates $(t,x,y,z)$ via
\begin{align}\label{bjork-transf}
    \tau=\sqrt{t^2-z^2}, \quad \tanh \eta = \frac{z}{t}.
\end{align}

The evolution of the hydrodynamic and superfluid variables we study will be entirely given by the proper time, $\tau.$ The equations of motion for the $U(1)$ scalar field are given by \eqref{Eq:EomSigma} and \eqref{Eq:EomSigma2}. Explicitly in the Bjorken background and denoting $\tau$-derivatives via a prime, the scalar equations of motion now take the form
\begin{align}\label{Eq:Bjorkenrho1}
 \cond'' + \frac{ \cond'}{\tau} + \lambda  \cond^3 + \mass(T- T_c)  \cond -  \cond {\psi' }^2 &= - {C_{\kappa_1}}T \cond', \\\label{Eq:Bjorkenpsi1}
(  \cond^2 \tau \psi')' &= - C_{\kappa_2}T^3 \tau \psi'.
\end{align}
where we have defined
\begin{align}
\kappa_1 = C_{\kappa_1} T, \quad \,\, \kappa_2 = C_{\kappa_2} T^3,
\end{align}
with $C_{\kappa_1}$ and $C_{\kappa_2}$ being dimensionless constants.
For the hydrodynamic variables, we will only consider the evolution of the shear tensor. A simple way to parameterize this is a diagonal form of $\pi^\mn$ \cite{Heller:2015dha}
\begin{align}
\pi^\mn =  {\rm diag}\left(0, \frac{\tilde{\pi}}{2},  \frac{\tilde{\pi}}{2}, -\frac{\tilde{\pi}}{\tau^2}\right).\end{align}
It is convenient to use the dimensionless pressure anisotropy $\chi := 3 \tilde{\pi}/4 T^4$. The conservation of the energy-momentum tensor \eqref{emt} provides the equation for the evolution of the temperature
\begin{eqnarray}\label{Eq:TEq1-bjork}
\frac{\tau T'}{T}  =\frac{1}{3} (\chi- 1)+\mass\frac{ \cond^2 + 2 \tau  \cond  \cond'}{8T^3}+ \frac{\tau}{4T^3}\left(C_{\kappa_1}{ \cond'}^2+C_{\kappa_2}T^2{\psi' }^2\right).
\end{eqnarray}
Finally, the MIS equation \eqref{mis-visc}, describing the evolution of the anisotropy, is given by
\begin{align}\label{Eq:ChiEq-bjork}
\tau \chi' + \frac{4}{3} \left(\chi - \frac{C_\eta}{C_{\tau_\pi}} \right)+4 \chi \frac{\tau T'}{T} + \frac{\tau}{C_{\tau_\pi}} \chi T = 0.
\end{align}
where using conformality, we have defined shear viscosity $\eta$ and $\tau_\pi$ as,
\begin{align}
    \eta = C_\eta s = \frac{4}{3} C_\eta T^3,  \quad \,\,\, \tau_\pi = C_{\tau_\pi} T^{-1},
\end{align}
with $ C_\eta$ and $C_{\tau_\pi}$ being dimensionless constant.
As such, we have a four-dimensional phase space given by $T$, $\chi$, $\psi'$, $ \cond$ and $ \cond'$. Since we set $\mu = 0$, the phase is given by a constant value and thus decouples from the dynamics.

\begin{figure}
\includegraphics[width=0.4\textwidth]{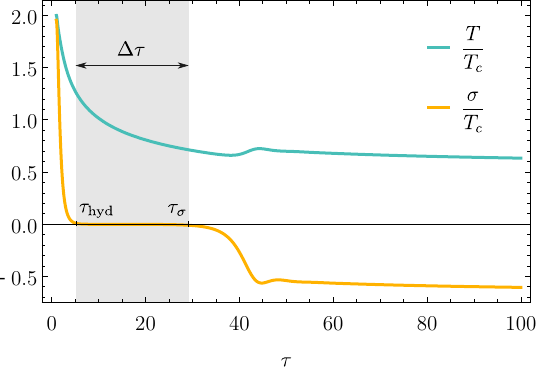}
\hspace{0.5cm}
\includegraphics[width=0.42\textwidth]{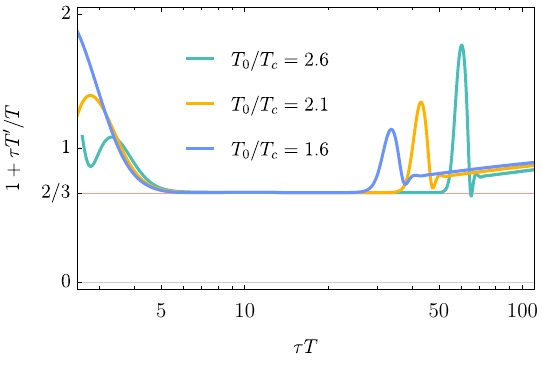}
 \caption{
 Left: Typical evolution for $T(\tau_0)>T_c$. The gray band denotes the length of time, $\Delta \tau,$ defined in \eqref{attractor-time}, when the system is predominantly following a hydrodynamic attractor behavior. The large $\tau$ behavior asymptotes to the gray dotted lines, namely the condensate tends to $-\sigma_*$ as in \eqref{condstar} and the temperature tends to $t_*$ as defined in \eqref{tempstar}. Initial conditions at $\tau=1$ correspond to $T_0/T_c = 2,$ $\cond_0/T_c=1.954$, $\cond^\prime=0$ and $\chi=1$.
    Right: Evolution of the rescaled dimensionless temperature $f = 1+\tau \, T'/T$ as a function of dimensionless time $\omega=\tau T$ for various initial values of temperature. For intermediate values of $\omega,$ the solutions approach the attractor whose asymptotic value is given by $2/3$ \cite{Heller:2015dha}. %This matches the \cite{Heller:2015dha} due to the fact that in that region the condensate vanishes. %The early-time and late-time behavior are richer due to a nonzero condensate. 
    Initial conditions at $\tau=1$ correspond to $\sigma_0/T_c = 3.5$, $\cond^\prime=0$ and $\chi=1$.}
    \label{fig:typ-bjork}
\end{figure}

We now briefly recap the results of \cite{Mitra:2020hbj} and then discuss the timescale between the two key regimes. 
The evolution of the superfluid in the Bjorken flow is characterized by the choice of initial conditions. We will primarily focus on the regime where $T>T_c$ initially. Initially, the potential in this case has one minimum, which the condensate quickly falls to, acquiring a value of $\cond=0.$ As such, the fluid no longer is influenced by the condensate and the dynamics of the system are that of the \textit{hydrodynamic attractor} (see \cite{Heller:2015dha}). The attractor solution is characterized by the temperature dropping like $T\sim \tau^{-1/3}$. Once the temperature falls below $T_c$, the shape of the potential changes. Eventually the condensate notices this and goes from the unstable local maximum at $\cond=0$ to one of the minima, $\cond=\pm\cond_*.$ This occurs non-monotonically with the temperature rising as the condensate oscillates around its new minimum. Finally, at late times, the system freezes, with the dissipative anisotropy, $\chi$, disappearing, the condensate attaining its final value of $\pm \sigma_*$ and the temperature reaching a non-zero final value, $t_*.$ That the system approaches a non-zero temperature at late times is a surprising feature of the present model, which is not found in typical Bjorken flow evolution in the absence of a condensate.

A typical evolution is shown in the left panel of Fig.~\ref{fig:typ-bjork}. In the broken phase, the condensate takes the following minimum at late times
\begin{align}\label{condstar}
    \cond_*=\sqrt{\frac{m_0 T_c(1-t_*)}{\lambda}},
\end{align}
where the asymptotic value of the temperature is
\begin{align}\label{tempstar}
    t_*= \lim_{\tau\rightarrow\infty}\frac{T(\tau)}{T_c}.
\end{align}
This final temperature to which the system approaches at late times is determined by the solution to the following equation
\begin{align}\label{eq:tstar}
    \frac{1-t_*}{t_*^3}=\frac{8\lambda T_c^2}{3m_0^2}.
\end{align}
We reiterate that \eqref{condstar} and \eqref{eq:tstar} are the late time ($\tau\rightarrow \infty$) static solutions for the condensate and the temperature, respectively, which follow from the equations of motion \eqref{Eq:Bjorkenrho1}, \eqref{Eq:TEq1-bjork} and \eqref{Eq:ChiEq-bjork}.

\begin{figure}

\includegraphics{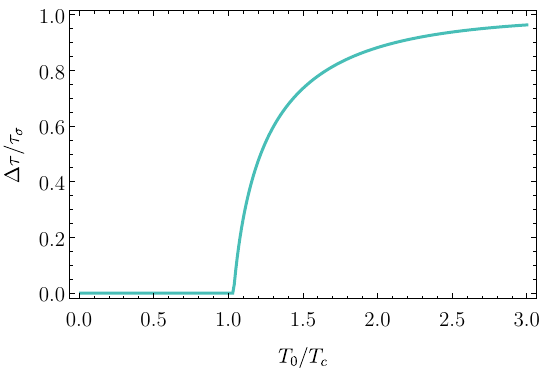}
    \caption{The normalized timescale to go from the onset of hydrodynamic attractor behavior, $t_{\rm hyd}$, to the onset of the condensate dominated regime, $t_{\cond}$, as a function of the initial temperature. The other parameters are mentioned in text. Essentially, this measures the region denoted by the gray band in the left panel of Fig.~\ref{fig:typ-bjork}.}
    \label{fig:bjorken}
\end{figure}

We define the timescale from the onset of hydrodynamic attractor behavior, $\tau_{\rm hyd}$, to the onset of the condensate regime, $\tau_{\cond}$, via
\begin{align}\label{attractor-time}
    \Delta \tau = \tau_{\cond} -\tau_{\rm hyd}.
\end{align} We note that $\tau_{\rm hyd}<\tau_{\rm \cond}.$ We can measure this interval of time by observing when the condensate remains close to zero
\begin{align}
\left\vert\cond(\tau)\right\vert\lesssim 10^{-2}.\label{tcond-condition}
\end{align}
This is the region where the condensate has no input in the dynamics of the viscous fluid, whose temperature goes like $T\sim \tau^{-1/3}$ in this region, see the right panel of Fig.~\ref{fig:typ-bjork}.\footnote{Another method to extract the timescale can be found by comparing the superfluid temperature, $T(\tau),$ to the temperature, $T_{\rm hyd}$, of a normal viscous fluid with a hydrodynamic attractor and finding the range of $\tau$ when 
\begin{align}\nonumber
    \left \vert T(\tau)-T_{\rm hyd}(\tau)\right\vert \lesssim 10^{-2}.
\end{align} 
We found that such a condition gave similar results, but for certain parameter ranges (especially for short $\Delta \tau$), the condition \eqref{tcond-condition} was more robust.}
In Fig.~\ref{fig:bjorken}, we show the generic duration of the attractor regime as a function of the initial temperature. We initialize at $\tau_0=1$ with $\cond=0.01$, $\cond^\prime=0$, $\chi=1$. For the scalar field, we choose $C_{\kappa_1}=1$, while for the fluid, we work with $C_{\eta}=1/4\pi$ and $C_{\tau_\pi}=(2-\log2)/2\pi,$ arising from matching MIS hydrodynamics to the holographic $N=4$ Super Yang-Mills theory \cite{Heller:2015dha}. We see that the higher the initial temperature, the longer the system remains trapped in the hydrodynamic attractor regime. In the limit of infinitely high temperature, $\tau_\cond \gg \tau_{\rm hyd}$ and the $\Delta \tau$ grows becomes infinitely large.

To interpret this timescale, it is helpful to think of a typical flow to describe heavy ion collisions. In the case of the normal Bjorken fluid, the time the system is well described by the hydrodynamic attractor is infinite. However, the physical system fails to have a hydrodynamic description as it freezes out and undergoes hadronization at some finite time. Hence, this model provides a picture of a fluid transitioning to a non-hydrodynamic description, which we can parameterize by $\Delta \tau$, the timescale which it takes for a Bjorken superfluid to escape the attractor regime.

We can make a back-of-the-envelope phenomenological estimate of this timescale. Using the following set of initial conditions at $\tau_0 = 1.4 \text{ fm}$ and parameters from \cite{Keegan_2016} and \cite{Grossi:2021gqi},
\begin{alignat}{4}
  &  m_0 = 198 \text{ MeV}, \qquad 
  && T_c = 155 \text{ MeV}, \qquad
  && C_{\eta} = 0.62,  \qquad
  && C_{\tau_\pi}=5.1 \times \frac{\pi}{2}   \nonumber \\ 
  &  T_{0} = 434 \text{ MeV}, \qquad
  &&\cond_{0} = 2 T_c, \qquad
  &&\chi_{0} = 0.1, \qquad
  && \cond'_{0} = 0.01 T_c  \label{bjork-estimate}
\end{alignat}
we find $\Delta \tau/\tau_0 \sim 13.2$. This is comparable to freeze-out, which is usually taken to be approximately in our units at $\tau/\tau_0 \sim 10$.

\section{Gubser Flow}\label{sec:gubs}
\noindent
Gubser flow is a time-dependent evolution of a many-body relativistic system, originally studied in the context of relativistic heavy-ion collisions \cite{Gubser:2010ze,Gubser:2010ui}. 
The flow describes a boost-invariant medium undergoing longitudinal and radial expansion while preserving the rotational symmetry. In a conformal system, the flow can be studied by mapping the Minkowski space $\mathbb{R}^{3,1}$ to a product of three-dimensional de Sitter space and the real line, $d S_3 \otimes \mathbb{R}$, which make the symmetry manifest. The explicit mapping to the $d S_3 \otimes \mathbb{R}$ space is done via Weyl rescaling $ds^2 \rightarrow  \tau^{-2}{ds^2}$ followed by a coordinate transformation of the Milne coordinates (see Fig.~\ref{fig:metrics}), which leads to the background
\begin{equation}
{\rm d}s^2_G = - {\rm d} \rho^2 + ~ L^2 \cosh^2 \left(\rho/L \right)({\rm d}\theta^2+\sin^2\theta {\rm d}\phi^2)  + L^2 {\rm d}\eta^2,
\end{equation}
where $(\rho, \theta)$ provides an alternate parameterization of the Milne coordinate $(\tau, x_\perp) $ by
\begin{align}\label{gubser-transf}
 \rho = - L\, \arcsinh{ \left( \frac{1-  q^2 \tau^2 +   q^2 x_\perp^2}{2 q \tau} \right)}, \quad 
 \theta = \arctan{\left(\frac{2 q x_\perp}{1+ q^2 \tau^2 - q^2 x_\perp^2}\right)},
\end{align}
where $q$ sets the transverse size of the colliding system (which we set to one), $\phi$ is the angular coordinate of the plane transverse to the $z$-axis, $\eta$ is the rapidity, and $L$ is the de Sitter length.

\subsection{Setup}
In this background, the scalar equations \eqref{Eq:EomSigma} are
\begin{align}\label{Eq:gubserrho1}
 \cond'' + 2 \tanh{(\rho/L)} \, \cond' + L^2\, \lambda  \cond^3 + L^2\, \mass(T- T_c)  \cond -  \cond {\psi' }^2 &= - L {C_{\kappa_1}}T \cond', \\\label{Eq:gubserpsi1}
(  \cond^2 \cosh^2{(\rho/L)} \, \psi')' &= - L \, C_{\kappa_2}T^3 \cosh^2{(\rho/L)} \, \psi',
\end{align}
where the prime denotes a derivative w.r.t.~the time $\rho/L$. 

We parameterize the dissipative tensor $\pi^\mn$ via the diagonal form  \cite{Denicol:2018pak}
\begin{align}
\pi^\mn =  {\rm diag}\left(0, \frac{\tilde{\pi}}{2 \cosh^2\left(\rho/L\right) L^2},  \frac{\tilde{\pi}}{2 \cosh^2\left(\rho/L\right) L^2 \sin^2\theta}, -\frac{\tilde{\pi}}{L^2}\right).
\end{align}
For convenience, we introduce the dimensionless pressure anisotropy $\chi := 3 \tilde{\pi}/4 T^4$.
We work in the local rest frame of the Gubser fluid, where we take the four-velocity to be $u^\mu = (1, \vec{0})$. Note that this generates a non-trivial flow if one were to undo the Weyl rescaling and coordinate transformation back to the Milne coordinates. The evolution equation of the temperature given by the conservation of the energy-momentum tensor \eqref{emt} reads
\begin{eqnarray}\label{Eq:TEq1}
\frac{ T'}{T}  = - \frac{\tanh \left(\rho/L\right)}{3} (\chi+2)
+\mass\frac{ \cond^2\tanh \left(\rho/L\right) +   \cond  \cond'}{4T^3}
+ \frac{1}{4 L T^3}\left(C_{\kappa_1}{ \cond'}^2+C_{\kappa_2}T^2{\psi' }^2\right).
\end{eqnarray}
and the MIS equation \eqref{mis-visc} is given by
\begin{align}\label{Eq:ChiEq}
\chi' + \frac{4 \tanh \left(\rho/L\right)}{3} \left(2\chi + \frac{C_\eta}{C_{\tau_\pi}} \right)+4 \chi \frac{ T'}{T} + \frac{L \chi T}{C_{\tau_\pi}}  = 0.
\end{align}

\begin{figure}
\includegraphics[width=0.4\textwidth]{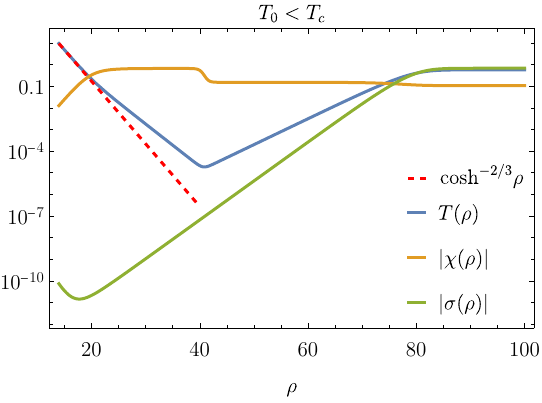}
\hspace{0.5cm}
\includegraphics[width=0.4\textwidth]{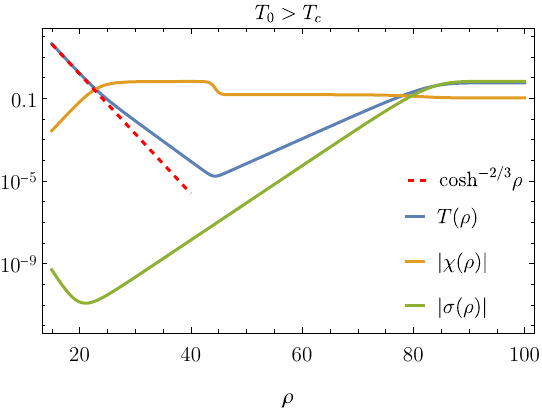}
    \caption{The figure shows temperature $T$, condensate $\sigma$ and pressure anisotropy $\chi$ for initial temperature $T_0<T_c$ and $T_0>T_c$ at initial de Sitter time $\rho_0 = -20$.  The left plot shows the evolution of the system for $T_0<T_c$ where $T_0 = 0.1 T_c$ and the right plot shows $T_0>T_c$ with $T_0 = 1.5 T_c$. The dashed red curve shows the hydro-like behavior of the system at some intermediate de Sitter time $\rho$. For both the plots, other initial conditions are set to $\cond_0 = 0.01$, $\chi(\rho_0) = 0.1$ and $\cond'(\rho_0) = 0$. The parameters used here are $\mass = 1$, $T_c = 1$, $C_{\tau_\pi} = (2-\log2)/2\pi$, $C_{\kappa_1} = 1$, $\lambda = 1$ and $C_{\eta}/C_{\tau_\pi} = 0.42$.}
    \label{Fig:gubser_solution}
\end{figure}

\begin{figure}
\includegraphics[width=0.5\textwidth]{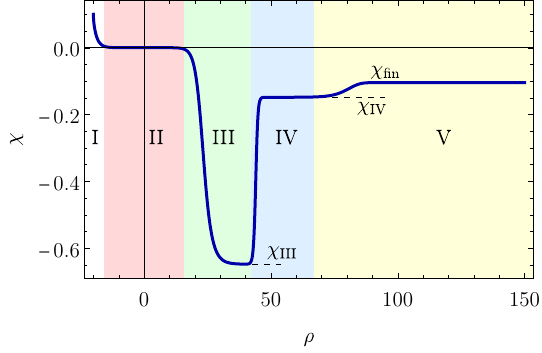}
    \caption{The evolution of the anisotropy, $\chi$, for $C_{\eta}/C_{\tau_\pi} = 0.42$ for $T_0>T_c$ with the same parameters as in Fig.~\ref{Fig:gubser_solution}. The system is initialized at $\rho_0 = -20$ with initial conditions given by $T_0 = 1.5 T_c$, $\cond_0 = 0.1$, $\chi_0 = 0.1$ and $\cond'_0 = 0$. We see that $\chi$ in the regions \textbf{III}, \textbf{IV} and \textbf{V} takes the value $\chi_{\rm III} \approx -0.648$, $\chi_{\rm IV} \approx -0.148$ and $\chi_{\rm fin} \approx -0.106$, respectively.}\label{Fig:gubser_chi}
    
\end{figure}

\begin{figure}
\includegraphics[width=0.45\textwidth]{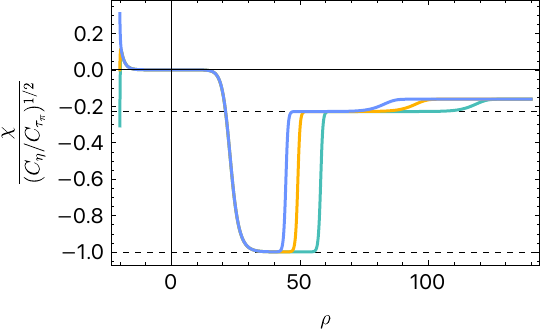}
\includegraphics[width=0.44\textwidth]{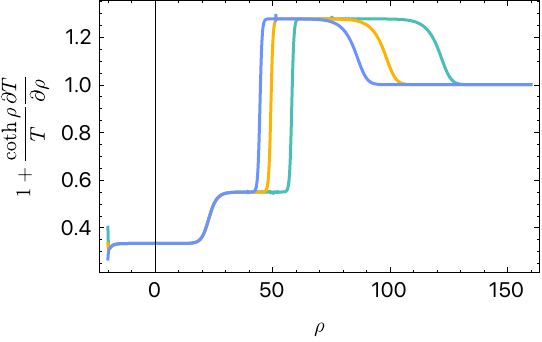}
    \caption{Left: Evolution of the anisotropy, $\chi/\sqrt{C_{\eta}/C_{\tau_\pi}},$ for different initial conditions of $\chi_0$ and fixed initial temperature, $T_0 = 1.2$,  and condensate, $\sigma_0 = 0.1$, with the same parameters as in Fig.~\ref{Fig:gubser_solution}. The system is initialized at $\rho_0 = -20$. The lines $\{0,-0.23,-1\}$ show that for arbitrary initial conditions the system approaches the attractor value given by $\chi_{\rm III}$, $\chi_{\rm IV}$, and region $II$. Right: The evolution of the system in terms of a dimensionless variable \cite{Behtash:2017wqg}. Note the evolution of the system out of the later regions depends on the initial conditions, as studied more closely in Figs.~\ref{fig:condplushydro_plot_Ck1eq4_tempfixed}-\ref{fig:condplushydro_plot_Ck1eq4_condfixed}.}\label{Fig:gubser_att}
    
\end{figure}

Using this set of equations, we study the evolution of the superfluid Gubser flow at a constant phase $\psi$ for arbitrary initial conditions. The remaining dynamical variables $T$, $ \cond$ and $\chi$, are function of de Sitter time $\rho/L$ only. Here we have set $L=1$ for computation.

\subsection{Results}
Unlike the Bjorken superfluid discussed above and in \cite{Mitra:2020hbj}, the evolution of the Gubser superfluid experiences similar evolution if the system is initialized in the unbroken phase $T_0>T_c$ or the broken phase $T_0<T_c$. In both phases, the system in the intermediate de Sitter time ($\rho = 0$) evolves to the hydro-like behaviour followed by the formation of condensate in the final state as shown in Fig.~\ref{Fig:gubser_solution}.
Moreover, we observe that the evolution of the system depends on the ratio $C_\eta/C_{\tau_\pi}$ for a fixed $C_{\kappa_1}$, dividing the evolution into four or five distinct regions, which we labelled \textbf{I-V} in Fig.~\ref{Fig:gubser_chi}. The richness of the dynamics in the Gubser superfluid flow is due in part to the interplay between the condensate and the anisotropy's evolution per the MIS formalism, which leads to a set of nonlinear equations \eqref{Eq:TEq1} and \eqref{Eq:ChiEq}. We further show in Fig.~\ref{Fig:gubser_att} that for a variety of initial conditions, the system enters an attractor regime for intermediate times, before exiting due to the effect of the condensate. This is analogous to the situation described in the superfluid Bjorken flow in the previous section.    

The region \textbf{I} represents the initial state, which is non-universal. The regions \textbf{II} and \textbf{V} are similar to the Bjorken case where \textbf{II} corresponds to inviscid hydrodynamics and \textbf{V} is the final state of the system when the condensate has been formed. The new regimes \textbf{III} and \textbf{IV} emerge due to the nonlinear set of differential equations and depend on the ratio of $C_\eta/C_{\tau_\pi}$. It is important to note that requiring a causal evolution sets an upper bound to the ratio $C_\eta/C_{\tau_\pi}$ \cite{romatschke2019relativistic} 
\begin{align}\label{ratio}
   \frac{ C_\eta}{C_{\tau_\pi}} \leq 1/2.
\end{align}
Furthermore, the condensate's dependence on the damping parameter $C_{\kappa_1}$ plays a crucial role in determining the timescale over which each region is spanned and the condensate's evolution to the final region \textbf{V}. We first discuss in details the behaviour of the system in each of the regimes for fixed $C_{\kappa_1}$. Afterwards, we will turn our attention to the consequences of $C_{\kappa_1}$ dependence on these regions and the condensate.

\begin{itemize}
    \item  \textbf{Region II} - \textbf{Perfect Hydro}
\par The region about $\rho = 0$ is when the system is dominated by perfect fluid-like behaviour. In this region, the condensate and the anisotropy term quickly approach zero, and the system is solely governed by \eqref{Eq:TEq1}. The temperature is given by
\begin{align}
    T_\text{ideal}(\rho) = {T_0}\cosh^{-2/3} \left(\rho/L \right),
\end{align}
where $T_0$ is a positive constant at some initial time.
    
    \item \textbf{Region III} - \textbf{Viscous hydro}
\par The region with vanishing condensate is characteristic of the conformal Israel-Stewart formalism of Gubser flow \cite{Nopoush:2014qba,Denicol:2018pak}. In this regime, the system exhibits high viscosity at a very low temperature. The equation that governs this region reads
\begin{align}
\chi' + \frac{4 \tanh{(\rho/L)}}{3} \frac{C_\eta}{C_{\tau_\pi}} - \frac{4}{3} \chi^2 \tanh{(\rho/L)} + \frac{L  \chi  T}{C_{\tau_\pi}}  = 0,\label{Eq:chi_regionIII}
\end{align}
which is obtained by observing that $\cond^2/T^3 \rightarrow 0$ and that although $T\rightarrow 0$, $T^\prime/T\neq 0$ and is determined by \eqref{Eq:TEq1}.  
We observe that the linear term in $\chi$ is almost negligible compared to the other terms due to the low temperature. Thus, the solution to this equation in the limit $\chi' \rightarrow 0$ and $\tanh{(\rho/L)} \rightarrow 1$ is given by
 \begin{align}
     \chi_{\rm III} = \pm {\left(\frac{C_\eta}{C_{\tau_\pi}}\right)}^{1/2}
 \end{align}
for arbitrary initial conditions, as shown in Fig.~\ref{Fig:gubser_chi}. Note however that this value may not be reached if there is significant dissipation in the equation of motion for the condensate, see the discussion in Sec.~\ref{sec:transitionIItoIV}.

\begin{figure}
\includegraphics[width=0.4\textwidth]{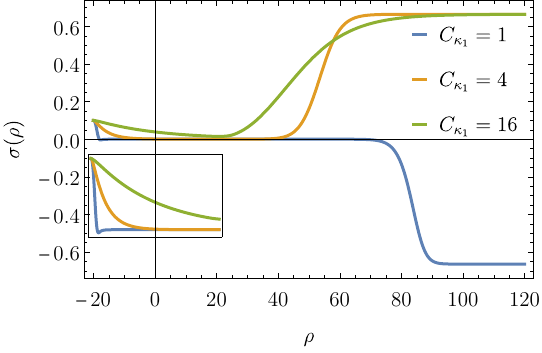}
\hspace{1cm}
\includegraphics[width=0.4\textwidth]{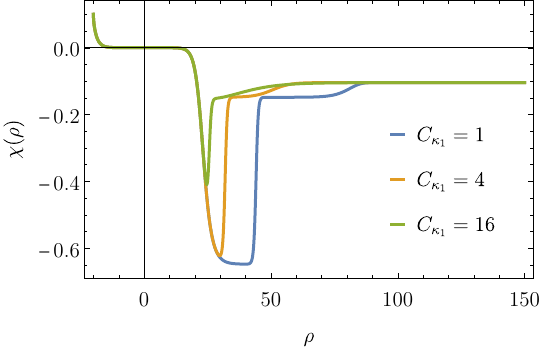}
\caption{The figure shows the evolution of condensate $\cond(\rho)$ and 
the anisotropy $\chi(\rho)$ for different $C_{\kappa_1}$ and fixed $C_\eta/C_{\tau_\pi} = 0.42$. The inset in the left figure shows the approach of the condensate to zero depending on $C_{\kappa_1}$. Here, we have chosen the initial condition for temperature and the condensate to be $T_0 = 1.5 \, T_c$ and $\cond_0 = 0.6 \, T_c$ by keeping the rest parameters the same as Fig.~\ref{Fig:gubser_solution}.}
    \label{fig:cond_plot_diffCk1}
\end{figure}

    \item \textbf{Region IV} - \textbf{Non-linear regime}
 \par  Despite the zero condensate value and almost negligible temperature, the system in the subsequent region shows a non-trivial evolution due to the non-vanishing ratio $\cond^2/T^3$, which otherwise was vanishing in the prior regions. We observe that in this region, other ratios that were previously vanishing become important, particularly $\cond'/\cond$.
 
 Hence, it is useful to rearrange the equations ($\tanh{(\rho/L)} $ has been set to one) to understand the system's behaviour associated with these ratios
\begin{align}\label{Eq:gubser_cond_regionIV}
  \left(\frac{\cond'}{\cond}\right)'+\frac{{\cond'}^2}{\cond^2} + 2 \frac{\cond'}{\cond} + L^2~\mass(T- T_c) + {C_{\kappa_1}}T ~L~ \frac{\cond'}{\cond} =0 ,\\
\label{Eq:gubser_temp_regionIV}
\frac{ T'}{T}  + \frac{1}{3} (\chi+2)
- \frac{\cond^2 }{4T^3} \left(\mass + \mass \frac{\cond'}{\cond}
+ C_{\kappa_1}\frac{{ \cond'}^2}{\cond^2} \right) =0,\\
\label{Eq:gubser_chi_regionIV} \frac{\chi'}{\chi} + \frac{4}{3} \left(2 + \frac{C_\eta}{C_{\tau_\pi} \chi} \right)+4  \frac{ T'}{T} + \frac{ L T}{C_{\tau_\pi}}  = 0.
\end{align}
We proceed with equation \eqref{Eq:gubser_cond_regionIV} and impose the limit $T \rightarrow 0$ which gives
\begin{align}
  \cond = e^{-\rho/L} \cosh{(\sqrt{1+L^2 ~ \mass T_c  }\rho)}.
\end{align}
We have set the integration constant to zero without any loss of generality. Next, using the non-zero contribution of the term $\cond^2/T^3$
we determine the constant ratio $T'/T$ for any $\rho$ value in this region 
\begin{align}
\frac{T'}{T} = -\frac{2}{3} \left(1 - \sqrt{\mass T_c L^2 + 1} \right) \equiv t_{p}.
\end{align}
Using this ratio further in equation \eqref{Eq:gubser_chi_regionIV}, we find the value of  $\chi$ to be
\begin{align}
\chi_{\rm IV} = -\frac{C_\eta}{C_{\tau_\pi}} \frac{1}{2 + 3 t_{p}}.
\end{align}
Note that we can also determine the value of $\cond^2/T^3$ from equation \eqref{Eq:gubser_temp_regionIV}
\begin{align}
  \left(  \cond^2/T^3 \right)_{\rm const} = \frac{4 L (3 t_{p}+ \chi_{\rm IV} +2)}{3 \left( C_{\kappa_1} \left(\frac{\cond'}{\cond}\right)^2+ L \mass (\frac{\cond'}{\cond} + 1) \right)}
\end{align}
for any value of $\rho$ in this region. 

It is important to highlight that although the condensate is extremely small in this region, the evolution of the anisotropy in this region depends not just on the fluid transport parameters but also on the parameters of the condensate, namely $\mass$.

\item  \textbf{Region V - Formation of condensate}\label{sec:gubser-cond}

This final region corresponds to the broken phase of the system, which is associated with the formation of the condensate, $\cond$. The system in this region attains a constant temperature and is characterized by a pair of symmetry-breaking fixed points similar to the Bjorken case \cite{Mitra:2020hbj}. However, in this case, the pressure anisotropy also attains a finite value, which otherwise vanishes in the Bjorken superfluid.  
In this region at large $\rho$, the derivatives go to zero, and we get the following set of algebraic equations
\begin{align}
 & \cond_{\rm fin}^2  =   \frac{\mass (T_c - T_{\rm fin} )}{\lambda}, \\
   &\chi_{\rm fin} = \frac{4 C_\eta}{3 C_{\tau_\pi}} \left(  \frac{L~ T_{\rm fin}}{C_{\tau_\pi}} + \frac{8}{3}  \right)^{-1},\\
& T_{\rm fin}^3 = \frac{3}{4} \frac{m_0^2 ( T_c - T_{\rm fin}  )}{(\chi_{\rm fin} + 2)\lambda},
\end{align}
which we can solve to find a unique set of solutions.

\end{itemize}

\begin{figure}[tbp]
\includegraphics[width=0.4\textwidth]{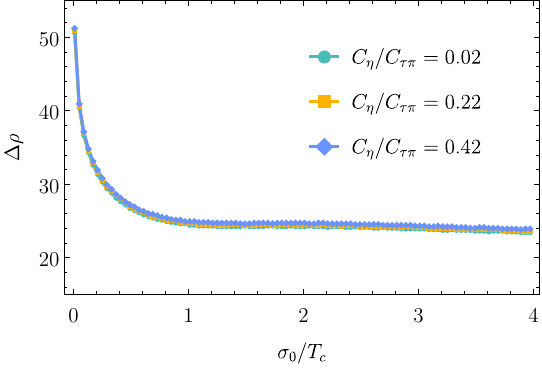}
\hspace{1cm}
\includegraphics[width=0.4\textwidth]{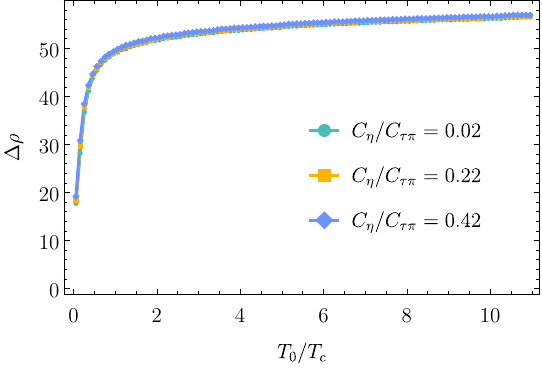}
  \caption{The behavior of the attractor time, $\Delta\rho$, as a function of initial conditions, namely $\sigma_0/T_c$ initially in the unbroken phase $T_0/T_c = 1.5$ (left) and as a function of $T_0/T_c$ with $\sigma_0/T_c = 0.01$ (right), is essentially independent of the ratio of hydrodynamic transport parameters $C_\eta/C_{\tau_\pi}$. Chosen parameters are $\lambda=4$ and $C_{\kappa_1} = 4$. 
  }
  \label{fig:condplushydro_plot_Ck1eq4_tempfixed}
\end{figure}

\begin{figure}
\includegraphics[width=0.4\textwidth]{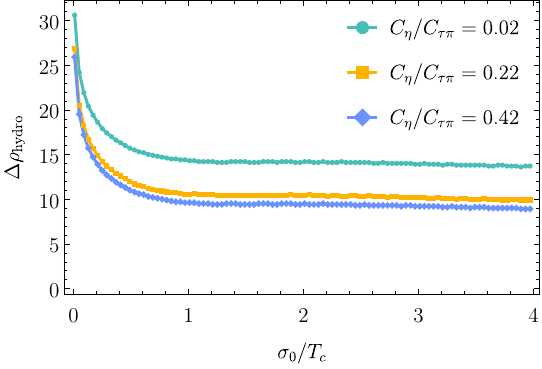}
\hspace{1cm}
\includegraphics[width=0.4\textwidth]{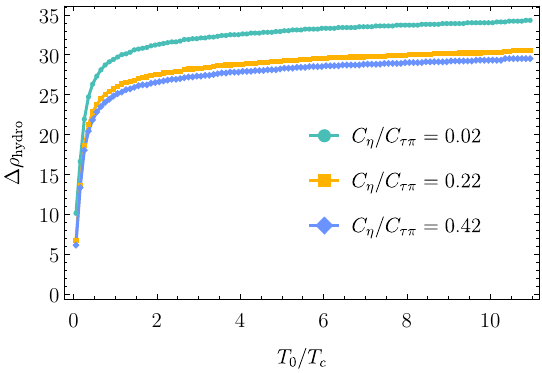}
\includegraphics[width=0.4\textwidth]{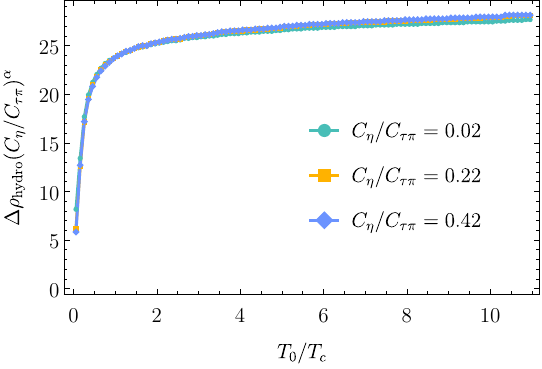}
  \caption{Behavior of $\Delta\rho_{\rm hydro}$ as a function of initial conditions, namely $\sigma_0/T_c$ (top left) and $T_0/T_c$ (top right). Unlike $\Delta\rho$ shown in Fig.~\ref{fig:condplushydro_plot_Ck1eq4_tempfixed} (see the figure caption for parameters used here), $\Delta\rho_{\rm hydro}$ depends on the ratio $C_\eta/C_{\tau_\pi}$. Bottom panel: rescaled $\Delta\rho_{\rm hydro}$ with the ratio $\left({C_\eta}/{C_{\tau_\pi}}\right)^{0.055}$ indicating a weak dependence on the ratio in this case. }
\label{fig:condplushydro_plot_Ck1eq4_tempfixed2}
\end{figure}

\subsection{Transition time from region \textbf{II} to region \textbf{V}}\label{sec:transitionIItoIV}
The above observation suggests a universal transition of the system from hydro-like behaviour to symmetry-breaking fixed points for any arbitrary initial conditions. Following the discussion in the Bjorken section, we can also make an estimate of the timescale of the Gubser hydro-like behaviour, which corresponds to region \textbf{II}, which we will denote as $\Delta \rho_{\rm hydro}$ and the transition time from region \textbf{II} to non-zero condensate region \textbf{V}, which we call $\Delta\rho$. However, as detailed above, due to the additional intermediate evolution, region \textbf{IV}, which is not present in the Bjorken case, we cannot make a direct comparison between $\Delta \rho$ and the $\Delta \tau$ defined in \eqref{attractor-time}.

We set the following requirements for the two timescales
\begin{align}
\cond(\Delta\rho) \sim 10^{-3}, \\
    \chi(\Delta\rho_{\rm hydro}) \sim 10^{-3}, 
\end{align}
where $\Delta\rho_{\rm hydro}$ corresponds to the duration over which $\chi = 0$, corresponding to the length of time the system is undergoing inviscid hydrodynamic evolution. Meanwhile, $\Delta\rho$ is a closer proxy to $\Delta \tau$ in \eqref{attractor-time}, as it indicates the length of time the evolution is not dominated by the condensate.

Typically, the duration of the transition from region \textbf{II} to region \textbf{V} not only depends on the initial conditions and the ratio $C_\eta / C_{\tau_\pi}$ but also on $C_{\kappa_1}$. As we observe in the left panel of Fig.~\ref{fig:cond_plot_diffCk1}, with increasing $C_{\kappa_1}$, the condensate decays slowly and takes zero value only for a short, intermediate time, thus influencing the duration of the region \textbf{III-IV}. This is evident in the right panel of Fig.~\ref{fig:cond_plot_diffCk1}, where the anisotropy never reaches $\chi_{\rm III}$ precisely because the condensate rolls so quickly to the bottom of the potential once the system enters the broken phase.

For $T_0 > T_c$, we note that the ratio of hydrodynamic transport coefficients $C_\eta/C_{\tau_\pi}$ has no effect on the length of time it takes for the system to reach the condensate regime $\Delta \rho$, see Fig.~\ref{fig:condplushydro_plot_Ck1eq4_tempfixed}. We see that $\Delta\rho$ decreases with increasing $\cond_0$ and saturates at large values of $\cond_0$. Similarly, for different $T_0$ at a fixed $\cond_0$, the transition timescale increases for large $T_0$. This is not the case for $\Delta \rho_{\rm hydro}$, which depends on the ratio. However, this is rather weak, as can be seen in the bottom panel of Fig.~\ref{fig:condplushydro_plot_Ck1eq4_tempfixed2}, where the $\Delta \rho_{\rm hydro}$ curves with a wide range of $C_\eta/C_{\tau_\pi}$ collapse into one when scaled appropriately by a power of $\left(C_\eta/C_{\tau_\pi}\right)^{0.055}$.

\begin{figure}
\includegraphics[width=0.4\textwidth]{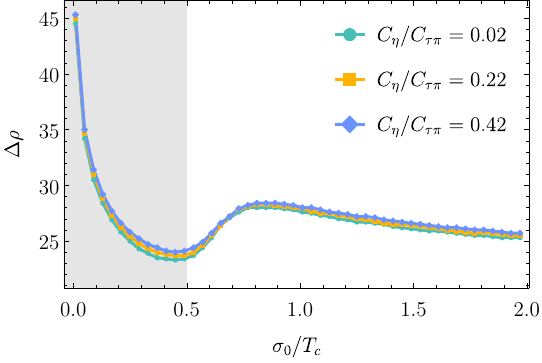}
\hspace{1cm}
\includegraphics[width=0.4\textwidth]{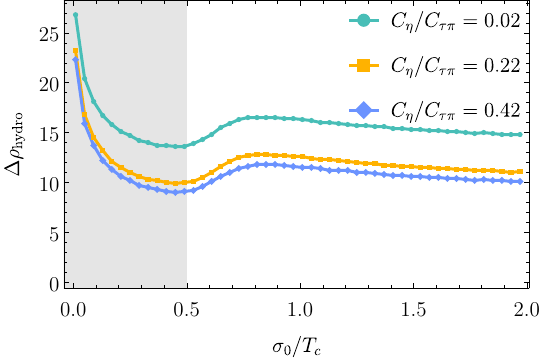}
\includegraphics[width=0.4\textwidth]{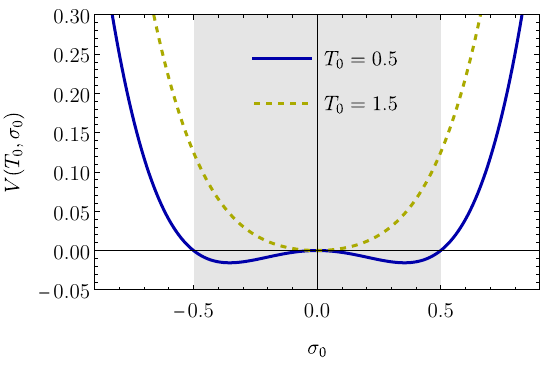}
 \caption{The top left and right figure show the $\Delta\rho$ and  $\Delta\rho_{\rm hydro}$, respectively, as a function of the initial conditions $\sigma_0/T_c$ for $T_0/T_c =0.5$ (other parameters are the same as in Fig.~\ref{fig:condplushydro_plot_Ck1eq4_tempfixed}). The bottom plot shows the potential as a function of initial condition $\cond_0$ and $T_0$.
 The grey shaded region marks when the condensate and the hydro timescale decrease before again increasing for this choice of parameters.
 }
\label{fig:condplushydro_plot_Ck1eq4_condfixed}
\end{figure}

So far, there have been little distinguishing features between initializing in the broken or unbroken phases for Gubser superfluid flow, compared to Bjorken flow. However, we see in Fig.~\ref{fig:condplushydro_plot_Ck1eq4_condfixed}, where $T_0 < T_c$, that unlike in the initialization in the unbroken phase seen in the left panels of Figs.~\ref{fig:condplushydro_plot_Ck1eq4_tempfixed} and \ref{fig:condplushydro_plot_Ck1eq4_tempfixed2}, the timescales have a non-monotonicity as a function of initial condensate. This can be explained by considering the potential, which is initially negative when $T_0 < T_c$ for a range of initial $\cond_0$ and finally changes sign when $V(\sigma_0, T_0) = 0$. This occurs whenever
\begin{align}
\vert\cond_0 \vert \leq  \sqrt{2 \frac{\mass (T_c-T_0)}{\lambda_0} } ,
\end{align}
as marked by grey in the bottom panel of Fig.~\ref{fig:condplushydro_plot_Ck1eq4_condfixed}. Initializing above this value then leads to agreement with the tendency for $T_0>T_c$.

Now we provide a phenomenological estimate of $\Delta \rho_{hydro}$ and $\Delta \rho$. We set the initial time to be $\rho_0 = -0.014$, which in Bjorken coordinates for unit transverse length $x$ and $q=1$ gives $\tau_0 = 1.4$ fm/c.  Similar to the Bjorken case \eqref{bjork-estimate}, we choose the initial conditions and parameters to be
\begin{alignat}{4}
  &  m_0 = 198 \text{MeV}, \qquad
  && T_c =155 \text{MeV},  \qquad
  && C_{\eta} = 1/4 \pi, \qquad
  && C_{\tau_\pi} = (2-\log2)/2\pi   \nonumber \\ 
  &  T_{0} = 434 \text{MeV}, 
  &&\cond_{0} = 0.1 T_c,  \qquad
  &&  \chi_{0} = 0.1,  \qquad
  && \cond'_{0} = 0.   \nonumber 
\end{alignat}
We also fix $L=1$. Note that in choosing the parameters, we have relaxed the assumption \eqref{ratio}, which, as a consequence, means that $\Delta \rho_{hydro} \sim \Delta \rho$. Thus, 
we find %$\Delta \rho_{hydro} \sim \Delta \rho\sim  2.5$, and also 
$\Delta \rho \sim 2.5$
which in Bjorken coordinates gives %$\Delta \tau_{hydro} \sim 12.3$ fm/c and also 
%$\Delta \tau_{\cond} \sim 12.3$ fm/c.
$\Delta \tau/\tau_0 \sim 8.8$.

\subsection{Gubser flow in physical background}

The mapping of the flat Minkowski spacetime, $(t,x,y,z)$ to the curved Gubser spacetime by Weyl rescaling manifests the symmetry of the flow, particularly the rotational symmetry, parameterized by the $(\theta, \phi)$ coordinates. This mapping simplifies the description of the fluid evolution by considering the fluid in a rest frame defined by the $dS_3 \otimes \mathbb{R}$ geometry. To visualize the flow in the physical background, one can undo the Weyl rescaling followed by appropriate coordinate transformations \eqref{gubser-transf}. Moreover, to compare the results obtained in  $dS_3 \otimes \mathbb{R}$ geometry to Bjorken spacetime, the physical quantities, such as the temperature, need to be appropriately rescaled after a Weyl transformation, as follows
\begin{align}\label{eq:backtoflat}
    T_{\rm B}(\tau, x_\perp) = \frac{1}{q \tau} T_{\rm G}(\rho/L). 
\end{align}
where $T_{\rm B}$ is the temperature in the flat spacetime.  Note that the fluid under consideration is conformal, therefore, the theory does not have any intrinsic scale, which makes $L$ the only scale in the background. As a consequence, the mapping of the physical observables from $dS_3 \otimes \mathbb{R}$ to $\mathbb{R}^{3,1}$ leads to the replacement of the arbitrary length scale $L$ by the transverse size $q$ of the colliding system. For further details of this transformation, the reader can refer to \cite{Banerjee:2023djb,Mitra:2024zfy}. 

\begin{figure}
\includegraphics[width=0.4\textwidth]{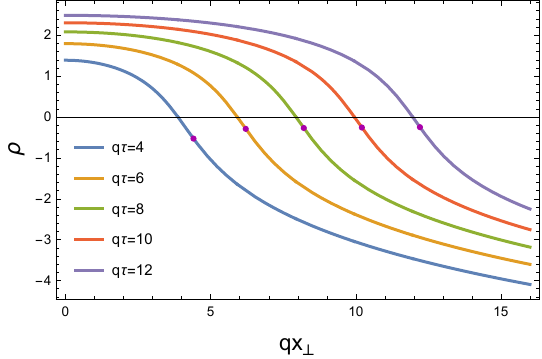}
\hspace{1cm}
\includegraphics[width=0.4\textwidth]{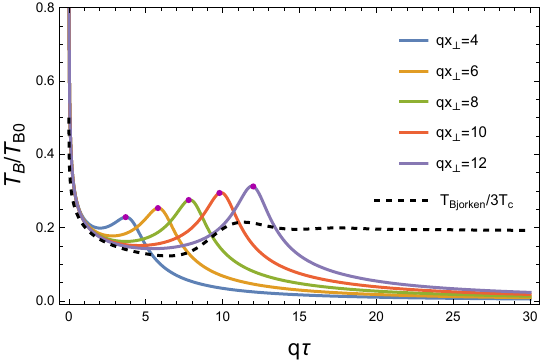}
 \caption{Mapping the Gubser solution in Fig.~\ref{Fig:gubser_solution} to Bjorken coordinates. The left panel shows $\rho$ as a function of the transverse coordinate $ qx_\perp$ for different choices of $q\tau$, where the magenta points are the same between the left and right panels. The top right panel shows the temperature after Weyl rescaling \eqref{eq:backtoflat}, normalized at $q\tau_0 = 0.01$ as a function of $(q\tau, qx_\perp)$. The black dashed line shows the evolution of the temperature in a Bjorken expansion for similar initial conditions.
 The initial conditions for both systems are the same, following the conditions in Fig.~\ref{Fig:gubser_solution}.}
\label{fig:Minkowski_gubser}
\end{figure}

 Figure \ref{fig:Minkowski_gubser} shows the Weyl-rescaled temperature as a function of \( (qx_\perp, q \tau) \). Initially, the temperature decreases, but at some intermediate time, it reheats, reaching a peak around \( qx_\perp \approx q\tau \) before eventually starting to decay. When comparing the evolution from the Gubser expansion in physical coordinates with those from the Bjorken expansion, illustrated by the black dashed line in Figure~\ref{fig:Minkowski_gubser}, we notice that both cases exhibit reheating features under similar initial conditions. However, in the Bjorken case, this reheating is associated to the dynamics of a non-zero condensate. In contrast, for the Gubser case, the expansion occurs in both the longitudinal and transverse directions, preventing the condensate from attaining a non-zero value within this period of proper time. As a result, the reheating observed in this range of proper time is solely related to the flow dynamics.

\section{FLRW}\label{sec:tldr}

\noindent
In this section, we turn our attention to the case of a dynamical metric. In previous sections, the Bjorken and Gubser backgrounds that we considered were fixed. While they let us track how a prescribed expansion rate drives the fluid-scalar system toward hydrodynamic attractors, they leave no room for the medium to backreact on the geometry. The natural next step is to promote the background to a dynamical degree of freedom. Cosmology provides a clear example of this: in a Friedmann-Lemaître-Robertson-Walker (FLRW) universe the expansion rate is a dynamical variable determined self-consistently through Einstein's equation. With that in mind, we now turn to the flat FLRW background
\begin{equation}
{\rm d}s^2_{\rm FLRW} = - {\rm d}t^2 + a(t)^2d\vec{x}^2,
\end{equation}
where $a(t)$ is the scale factor (see Fig.~\ref{fig:metrics} for relationship to the other metrics discussed here). We implement Einstein's equation by promoting our effective action \eqref{Eq:EffectiveAction} into the matter part $S_m$ of the Einstein-Hilbert action
\begin{equation}
    S_{\text{EH}} = \frac{1}{2\kappa} \int d^4 x \, \sqrt{-g} \, (R-2\Lambda) + S_m,
\end{equation}
where $\Lambda$ is the cosmological constant, $R$ is the Ricci scalar and $\kappa = {8 \pi G}/{c^4}$ is the gravitational constant (which we will set $\kappa=1$).
Varying the action leads to Einstein's equation,
\begin{equation}\label{eq:efe}
    R_{\mu\nu} - \frac{1}{2} R\, g_{\mu\nu} + \Lambda g_{\mu\nu} = \kappa \, T_{\mu\nu},
\end{equation}
where $T^{\mu\nu}$ is given by \eqref{emt}. We have one new dynamical parameter---the scale factor---for which we presumably need one additional equation. We decide to use the $(00)$ component of Einstein's equation \eqref{eq:efe}, which gives us the familiar Friedman equation, {in the fluid's local rest frame} 
\begin{equation}\label{eq:hubble}
    H^2 = \frac{\kappa}{3} T^{00}%E
    + \frac{\Lambda}{3},
\end{equation}
where $H = \dot{a}/a$ is the Hubble parameter. We will refer to this as the Hubble equation of motion.\footnote{We note that the other Friedmann equation is included automatically in the conservation of the energy-momentum tensor, which we take as a dynamical equation.}

\subsection{Setup}

\noindent
In this section, we will work with the energy density instead of temperature as we will need to include bulk viscous effects by including the bulk MIS equation \eqref{mis-visc-bulk}. This is due to the fact that the shear tensor vanishes in the FLRW background. Thus, the usual assumption of conformality that we used for the normal fluid in Bjorken and Gubser flow no longer holds in FLRW.\footnote{Note that in the Bjorken and Gubser case, the energy density is related to the temperature via $\varepsilon \sim T^4$ in four dimensions.} 
Hence, we are interested in general equations of states for the fluid sector of the form
\begin{equation}\label{eq:flrw_equation_of_state}
    p = w \, \varepsilon,
\end{equation}
where $w$ is a constant. Examples of typically studied equations of state include: $w = 0$ for matter, $w = 1/3$ for radiation (as considered in Sec.~\ref{sec:bjork} and Sec.~\ref{sec:gubs}), $w = -1$ for dark energy, and $w = 1$ for stiff matter \cite{chavanis2015CosmologyStiffMatter}.

Now we consider the equations of motion. Setting $\psi' = 0$ from the outset, the condensate evolution in the FLRW flow, given by \eqref{Eq:EomSigma}, is
\begin{align}\label{eq:flrw_condensate_eom}
    \sigma'' + 3H\sigma' + \lambda \sigma^3 + m_0 ( \varepsilon^{1/4} - T_c)\sigma = -C_{\kappa_1} \, \varepsilon^{1/4}  \sigma'
 \end{align}
where the prime denotes a derivative w.r.t.~the time $t$. 

Finally, we work with the parametrizations
\begin{equation}\label{eq:flrw_parametrizations}
    \tau_\Pi = C_{\tau_\pi} \, \varepsilon^{- 1/4}, \;\;\;\;\; \zeta = C_\zeta \, \varepsilon^{3/4}.
\end{equation}
where $C_{\tau_\pi}$ and $C_\zeta$ are dimensionless constants. 
The evolution of the energy density is given by the conservation of the energy-momentum tensor \eqref{emt}
\begin{align}\label{eq:flrw_energydensity_eom}
 \frac{\epsilon'}{\epsilon} 
    &+ 3H(1+w) (1+ \chi)
    - \frac{\mzero (1+w)}{8 w} \frac{3 H \sigma^2 + 2 \sigma \sigma'}{\epsilon^{3/4}} \nonumber\\
    &+ \frac{\mzero (3w-1)}{32 w} \frac{\sigma^2 \epsilon'}{\epsilon^{7/4}} - \Ckappasigma \frac{\sigma'^2}{\epsilon^{3/4}} = 0.
\end{align}
The MIS equation \eqref{mis-visc}, dictating the evolution of the {dimensionless bulk dissipation variable} $\chi=\Pi/(\varepsilon(1+w))$, is given by
\begin{equation}\label{eq:flrw_anisotropy_eom}
    (1+w) \chi' + (1+w) \left(\frac{\epsilon^{1/4}}{\Ctaupi} + \frac{\epsilon'}{\epsilon}       \right) \chi + 3 \frac{\Czeta}{\Ctaupi} H = 0.
\end{equation}
Finally, writing out the Hubble parameter evolution \eqref{eq:hubble}, we have our last equation
\begin{equation}\label{eq:flrw_hubble_eom}
    H^2 = \frac{\Lambda}{3} + \frac{\kappa}{3} \left( \varepsilon + m_0 \, \frac{\varepsilon^{{1}/{4}} - T_c}{2}  \sigma^2 - \frac{m_0 \, (w+1)}{8 \, w} \varepsilon^{1/4} \sigma^2 + \frac{1}{4} \lambda \sigma^4 + \frac{1}{2} \sigma'^2 \right).
\end{equation}

Upon inspection it can be seen that for $w = 0$ and $w = -1$ our equations reduce trivially. For the case of $w = 0$, the Hubble equation of motion constrains either $\sigma = 0$ or $\varepsilon =0$ both of which, in turn, leave us with only one evolution equation for multiple variables thereby making our system of equations underdetermined. Similarly, when $w = -1$ the MIS equation constrains either $H = 0$ or $\varepsilon = 0$. The former leaves us with no evolution equation for $H$ and the latter leaves us with no evolution equation for $\varepsilon$. Thus, we will focus our attention to the cases $w = 1/3$ and $w = 1$.

\subsection{Results}

The evolution of the system in the FLRW background can be characterised via three distinct regions, \textbf{I-III}, depending on the initial condensate $\cond_0$ at time $t_0 = 0$ and the ratio $C_\zeta/C_{\tau_\pi}$. The first region $\textbf{I}$ is the usual non-universal initial condition-dependent regime at time $t_0$. Region $\textbf{II}$ is characterised by the existence of an attractor-like behaviour when $\chi$ tends to $-1$. The approach of $\chi$ to this limit depends on $\cond_0$ along with the ratio $C_\zeta/C_{\tau_\pi}$. The final region $\textbf{III}$ is associated with the symmetry breaking fixed points similar to the Bjorken and Gubser cases. 

To develop some intuition before studying the complete FLRW evolution with a condensate, we will warm up by studying the following in a FLRW background: inviscid hydrodynamics, viscous hydrodynamics and a perfect fluid with a condensate. For concreteness, unless otherwise stated, the parameters we work with are
\begin{align}\label{flrw-parameters}
    w=1/3, \quad C_{\kappa_1} =\Lambda=T_c=m_0=\kappa= 1.
\end{align}
\begin{itemize}

  \item \textbf{Inviscid hydrodynamic regime:} $\sigma = 0$ and $\chi = 0$.
  
We begin by setting condensate and $\chi$ to zero, which corresponds to the FLRW perfect fluid. In this scenario, the energy density and Hubble parameter obtained from equations \eqref{eq:flrw_energydensity_eom} and \eqref{eq:flrw_hubble_eom} are given by 
\begin{align}\label{eq:flrw_energydensity_hubble_ideal}
 &   \varepsilon(t)_\ideal = - \frac{\Lambda}{\kappa}  \text{sech}^2\left\{\frac{1}{2} \sqrt{\Lambda } \left[\sqrt{3}  (w+1)t \right]\right\}, \nonumber \\
  &  H_\pm(t)_\ideal =  \pm \sqrt{\frac{\Lambda}{3}} \tanh \left\{\frac{1}{2} \sqrt{\Lambda } \left[\sqrt{3}  (w+1)t \right]\right\},
\end{align}
respectively. The integration constant is set to zero without any loss of generality. We see that at late times $H_\pm(t)$ approaches 
\begin{align}
    \lim_{t\rightarrow \infty}H_\pm(t)=\pm\sqrt{\frac{\Lambda}{3}},
\end{align}
while $\epsilon(t)$ approaches zero. This region corresponds to the perfect fluid-like behaviour of the flow where one can express the energy density in terms of scale factor $a(t)$
\begin{align}
    \varepsilon(t)_\ideal = \varepsilon_0 \, a(t)_\ideal^{-3 (w+1)}.
   \end{align}
In the limit of late times, $a(t)$ is given by
\begin{align}
    a(t) = a_0 e^{H_\const^\pm t},
\end{align}
and the corresponding energy density reads as
\begin{align}\label{eq_flrw_energy_density_vs_scale_factor}
    \epsilon(t)_\ideal = \epsilon_0 e^{-3(1+w) H_\const^+ t}.
\end{align}

   \item \textbf{Viscous hydrodynamic:} $\sigma = 0$ but $\chi \neq 0$.

In the viscous hydrodynamic case, we observe that for arbitrary initial conditions, $\chi(t)$ always approaches $-1$ for late times, irrespective of the parameter choice, while the energy density $\varepsilon(t)$ and the Hubble parameter $H(t)$ approach a constant value depending on the choice of parameters as shown in Fig.~\ref{Fig:flrw_anisotropy_to_-3}. In this case the late-time region is characterised by the following set of solutions obtained from \eqref{eq:flrw_energydensity_eom}-\eqref{eq:flrw_hubble_eom}
\begin{align}
&\chi = -1, \\
   &  \epsilon = -\frac{\Lambda }{\kappa } + (w+1)^2\frac{ (w+1)^2 + \sqrt{(w+1)^4-36 \kappa  \Lambda  \Czeta^4 }}{2 \left(3 \Czeta^2 \kappa \right)^2},\label{Eq:flrw_eviscous} \\
   &  H_{\pm} = (w+1)  \frac{\sqrt{(w+1)^2 + \sqrt{(w+1)^4-36 C_\zeta ^4 \kappa  \Lambda }}}{3 C_\zeta^2 \sqrt{6 \kappa }} \label{Eq:flrw_hviscous}.
\end{align}
\begin{figure}
\includegraphics[width=0.45\textwidth]{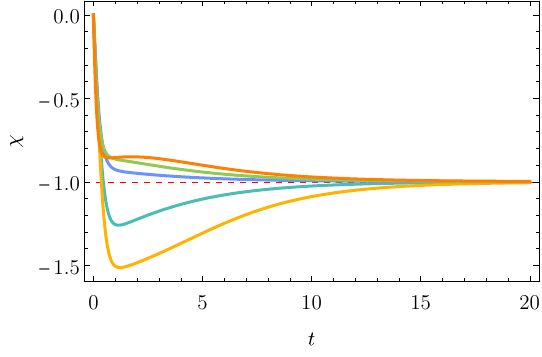}
\hspace{1cm}
\includegraphics[width=0.45\textwidth]{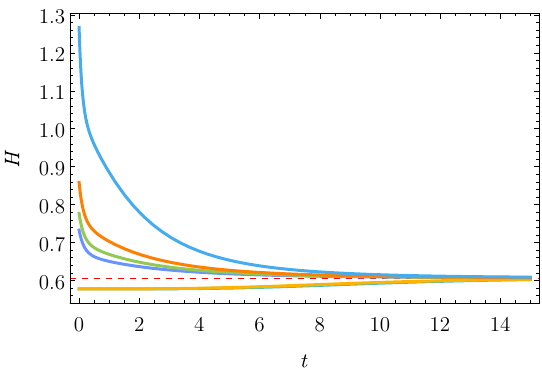}
    \caption{The dissipative fluid in FLRW with $\cond = 0$. Left: the dimensionless dissipative parameter, $\chi$, always converges to $\chi = -1$ at late times given by the red dashed line for arbitrary initial conditions and parameters. Right: The Hubble parameter $H(t)$ approaches a distinct final value depending on $w$, for fixed $\Lambda$ and $C_\zeta/C_{\tau_\pi}$. For the choice of parameters, see \eqref{flrw-parameters} and $C_\zeta/C_{\tau_\pi} = 2$.
    }
\label{Fig:flrw_anisotropy_to_-3}
\end{figure} 
   
    \item \textbf{Perfect fluid with condensate:} $\sigma \neq 0, \chi = 0$.
    
    Here we consider a perfect fluid coupled to a dynamical condensate. The small but finite initial condensate leads to three distinct regions, where the first region is the usual initial condition-dependent region. The second region is associated with the perfect fluid regime, following the solution \eqref{eq:flrw_energydensity_hubble_ideal}. Characteristically, the energy density is vanishing, while the Hubble parameter tends to $\sqrt{\Lambda/3}.$ Curiously, this holds even for larger values of the condensate, as can be seen in the left panel of Fig.~\ref{Fig:flrw_inviscid}. However, initializing with large values of the condensate means that the system never has the chance to undergo perfect fluid evolution, as shown in the right panel of Fig.~\ref{Fig:flrw_inviscid}.
    
    The final region is when the system evolves to one of the symmetry-breaking fixed points given by the solution to 
    \begin{align}
    &\varepsilon_\text{fin}^{3/4} = \frac{\, m_0^2 (T_c-\varepsilon_\text{fin}^{1/4})}{8 \, w \lambda  }, \\
 &   \sigma _{\text{fin}}^2 = \frac{m_0 \, (T_c - \varepsilon _{\text{fin}}^{1/4})}{\lambda }, \label{Eq:flrw_final_c} \\
   & H_\text{fin}^2 =  \frac{ m_0^2 \kappa (\varepsilon_\text{fin}^{1/4}-T_c)}{12 \lambda} \left(\frac{(w+1) \varepsilon_\text{fin}^{1/4}}{2 w } - (\varepsilon_\text{fin}^{1/4}-T_c)  \right) + \frac{\Lambda }{3} +\frac{\kappa  }{3} \varepsilon _{\text{fin}} .\label{Eq:flrw_final_h}
\end{align} 
Note that the above system of equations is solvable in closed form. However, as the solution is not particularly illuminating, we leave the above equations as they are.

\begin{figure}
\includegraphics[width=0.45\textwidth]{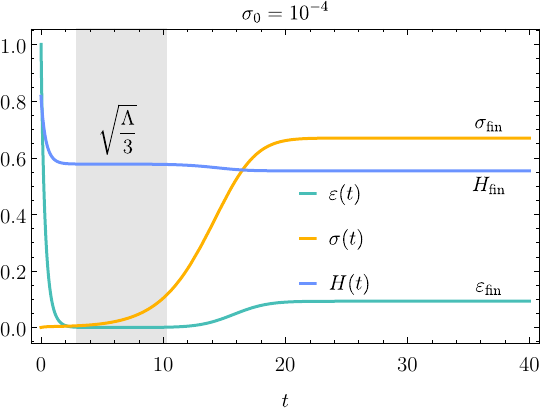}
\hspace{0.5cm}
\includegraphics[width=0.45\textwidth]
{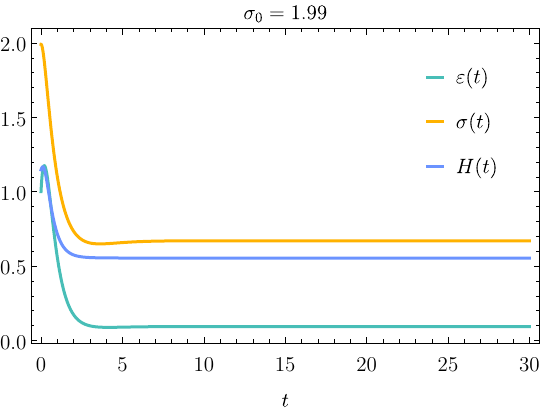}
    \caption{The evolution of condensate $\cond(t)$, energy density $\varepsilon(t)$, the Hubble parameter $H(t)$ for inviscid case for two different initial conditions of the condensate with parameters given by \eqref{flrw-parameters}. Left: sufficiently small values of the condensate lead to an intermediate regime described by the perfect fluid, given by \eqref{eq:flrw_energydensity_hubble_ideal}. Right: for larger initial values of the condensate, the system is dominated by the condensate and has no intermediate perfect fluid-like evolution.}
    \label{Fig:flrw_inviscid}
\end{figure} 

 \item \textbf{Full system:} $\sigma \neq 0$ and $\chi \neq 0$.

We now turn our attention to the full system.
We observe that for small but finite initial condensate, as shown in the top left panel of Fig.~\ref{fig:flrw_region}, the system in the region \textbf{II} behaves predominantly like in the viscous hydrodynamic scenario where $\chi=-1$ rather than the perfect fluid-like behaviour observed in the case of Bjorken and Gubser flow. In this region, the energy density $\varepsilon(t)$ and the Hubble parameter $H(t)$ follow the expression \eqref{Eq:flrw_eviscous} and \eqref{Eq:flrw_hviscous}. 

However, as mentioned above, for large initial condensate and low viscosity, before the system reaches the attractor, it evolves to the condensate-dominated region of symmetry-breaking fixed points whose equations are given by
\begin{align}
    &\varepsilon_\text{fin}^{3/4} = \frac{ \, m_0^2 (T_c - \varepsilon_\text{fin}^{1/4})}{8 \, w \lambda (1 + \chi_\text{fin})},  \label{Eq:flrw_final_e}\\
    & \chi _{\text{fin}} = -\frac{3 \, C_\zeta  H_\text{fin} }{(w+ 1) \varepsilon_{\text{fin}}^{1/4}},  \label{Eq:flrw_final_chi}
\end{align}
and the final values for the condensate and the Hubble parameter are the same as in the perfect fluid plus condensate \eqref{Eq:flrw_final_h} and \eqref{Eq:flrw_final_c}. Note that there are no closed form solutions to the above system of equations. It should be noted that, subject to the condition that $\varepsilon_{\rm fin}>0$ and $\sigma_{\rm fin},\, \chi_{\rm fin},\,H_{\rm fin}\in \mathbb{R}$, there exists a unique solution. 

Another visualization of the dynamics can be seen in the phase space plot Fig.~\ref{fig:flrw_phase_space}, where the energy density is plotted parametrically against the condensate. We see that the energy falls dramatically and quickly approaches its minimal value, when $\chi=-1.$ Moreover, irrespective of the choice of $\cond_0$, the condensate inexorably evolves to larger values, never decreasing to zero, which would indicate the onset of the attractor regime. 
\end{itemize}

\begin{figure}
\includegraphics[width=0.45\textwidth]{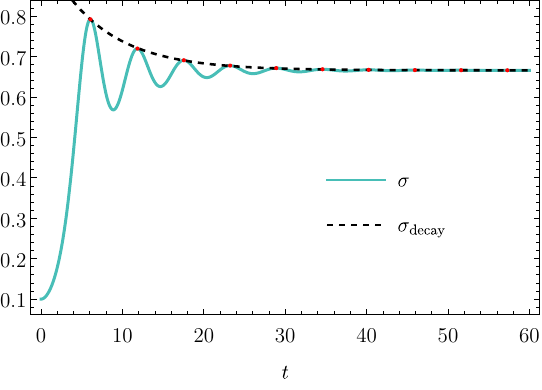}
\hspace{1cm}
\includegraphics[width=0.46\textwidth]{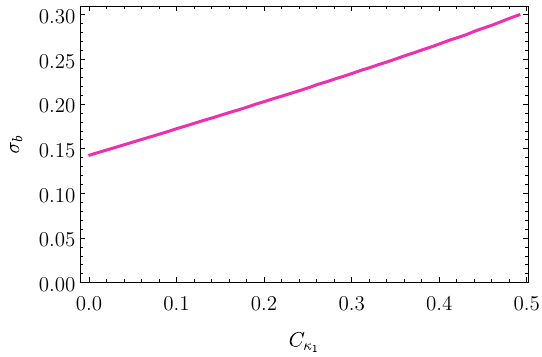}
    \caption{Left: The decaying amplitude $\sigma_\text{decay}$ of a typical solution for the condensate $\cond$ given a small $C_{\kappa_1}$, with the initial conditions $\varepsilon_0 = 0.1$, $\sigma_0 = 0.1$, $\sigma_0' = 0$ and $\chi_0 = 0$. The dotted line represents the fit \eqref{osc-fit} through the local maxima which are denoted with red points. Here we have taken the parameters \eqref{flrw-parameters} with $\Czeta = 1/10$, and $C_{\kappa_1}=0.01$. Right: The behaviour of the decay parameter $\sigma_b$ (see \eqref{osc-fit}) which increases as a function of $C_{\kappa_1}$.}
    \label{fig:flrw_condensate_decay}
\end{figure}

Finally, we comment on smaller values of dissipative transport coefficients. In this case, the condensate does not roll to the bottom of the potential, but instead overshoots it due to the smaller amount of friction. As can be seen in the left panel of Fig.~\ref{fig:flrw_condensate_decay}, the final value of the condensate, $\cond_{\rm fin}$, is approached via a decaying, oscillatory manner. Taking the local maxima of the oscillatory part of the $\cond$ solution, which we denote as red points in the left panel of Fig.~\ref{fig:flrw_condensate_decay}, we can fit the exponential decay to $\cond_{\rm fin}$ by
\begin{align}\label{osc-fit}
    \sigma_\text{decay} = \sigma_\text{fin}+ \sigma_a \, e^{- \sigma_b \, t}.
\end{align}
We see in the right panel of Fig.~\ref{fig:flrw_condensate_decay}  that the decay to the final value of the condensate is controlled by the friction coefficient $C_{\kappa_1}$, essentially with $\sigma_b\propto C_{\kappa_1}$ for small  $C_{\kappa_1}$.

\begin{figure}
\includegraphics[width=0.46\textwidth]{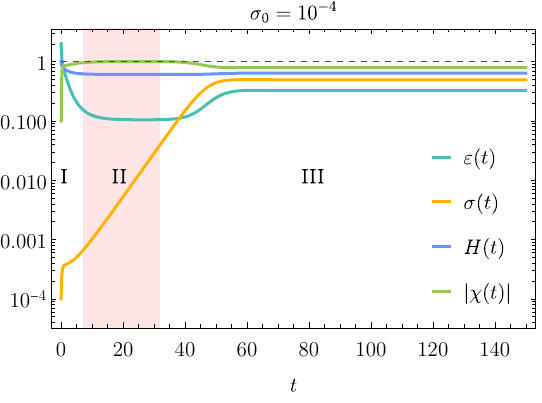}
\hspace{0.5cm}
\includegraphics[width=0.45\textwidth]{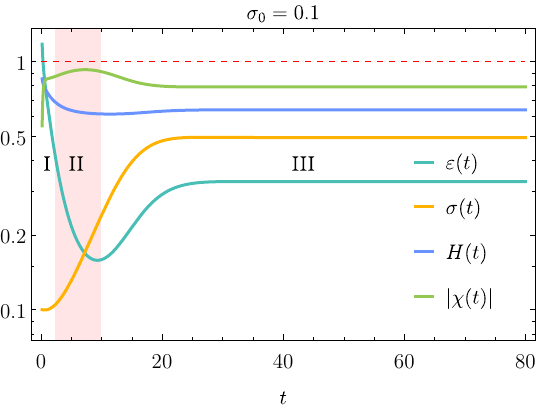}
\includegraphics[width=0.45\textwidth]{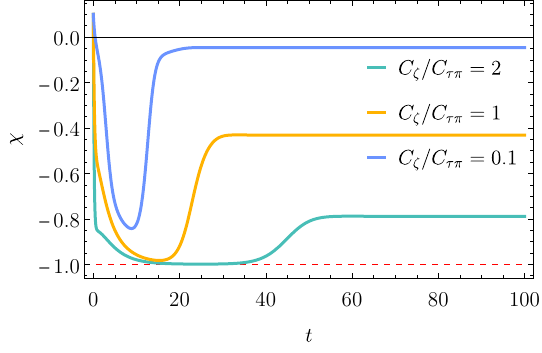}
    \caption{The evolution of the full system in the FLRW background. The top two plots show the evolution of the full system for different initial condensates $\cond_0$ at fixed $\epsilon_0 = 2$ and $C_\zeta/C_{\tau_\pi} = 2$. The red-shaded region marks the second region \textbf{II} where the system approaches the attractor given by the dashed red line $|\chi|=1$. The bottom plot shows the $\chi$ behaviour for different ratios $C_\zeta/C_{\tau_\pi}$ and fixed initial condensate, $\cond_0 = 10^{-4}$ and energy density $\epsilon_0 = 2$. The choice of parameters we work with is given by \eqref{flrw-parameters}.}
    \label{fig:flrw_region}
\end{figure}

\begin{figure}
\includegraphics[width=0.45\textwidth]{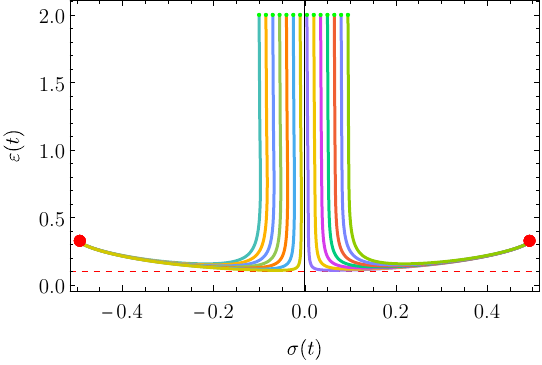}
    \caption{Phase space plot in the FLRW background. The initial conditions are such that $\varepsilon_0 = 2$ and $\chi_=0.1$, with parameters given by \eqref{flrw-parameters} and $C_\zeta/C_{\tau_\pi} = 2$. The green dots denote the initial conditions. The red dashed line shows the corresponding value of energy density, $\varepsilon = 0.1039$ when $\chi = -1$. The red dots correspond to the final fixed points of the system, $\cond_\pm=\pm 0.4934$. }
    \label{fig:flrw_phase_space}
\end{figure}

\section{Discussion}

In this work, we study superfluids, consisting of a normal fluid component and a scalar field that undergoes spontaneous symmetry breaking in a variety of expanding backgrounds. We primarily focused on spacetimes of particular interest to high energy particle physics, namely 
the boost-invariant Bjorken flow and the radially expanding boost-invariant Gubser flow. It is key to point out that in these spacetimes, we did not require any slow roll assumption: for a large class of arbitrary initial conditions (essentially $T>T_c$), the system would evolve following a certain generic behavior. Namely, this would involve a hydrodynamic-like evolution as the condensate fell into its minimum in the unbroken phase and a late time regime characterized by the condensate rolling into the new minimum in the broken phase. Furthermore, noting the similarities between our discussion of expanding systems undergoing a phase transition, we turned our attention to a cosmological model of a phase transition, namely FLRW with a scalar field. 

We defined the notion of attractor time, the length of time a system is trapped by the hydrodynamic attractor, before the condensate falls into a minimum in the broken phase at late times and studied its duration by varying the initial conditions. We found that in the Bjorken and Gubser case, the attractor time grew as a function of the initial temperature. Furthermore, we see that in the FLRW case, the existence of an attractor regime depended heavily on the initial value of the condensate. 

Moreover, we provided the first complete picture of a superfluid Gubser flow. The evolution differed significantly from the superfluid Bjorken flow, which for $T_0>T_c$ would approach the attractor regime at intermediate times before domination by the condensate at late times. The Gubser superfluid evolution for an appropriate choice of parameters from initial conditions first began with a regime dominated by perfect fluid hydrodynamics around $\rho=0$ with vanishing anisotropy until viscous MIS effects became important. In this regime, the anisotropy took a value related to the ratio of hydrodynamic transport coefficients, namely $(C_\eta/C_{\tau_\pi})^{1/2}.$ Following the viscous hydrodynamic evolution, we characterized a unique intermediate stage, arising from the nonlinearities of the equations, where although the temperature and condensate were small enough to potentially be negligible, the anisotropy took a constant value proportional to the ratio of hydrodynamic transport coefficients, $C_\eta/C_{\tau_\pi}.$ Finally, at late times, the dynamics become dominated by the condensate. Thus, we are able to quantify the asymptotic values that the system tends to, which can be expressed entirely in terms of the viscous transport coefficients.

There have been a number of simplifications in the present work to make a tractable model, which in subsequent works can be relaxed. For example, the dynamics presented here are completely classical: after the condensate falls into a minimum, it remains there for all time. In the present work, there are no false vacuums nor is there tunneling. A fully quantum treatment would require the inclusion of tunneling \cite{Matteini:2024xvg}, which is outside the scope of this work. 
Moreover, although in the class of models considered here there is a stage of the evolution where the system temperature increases which bears resemblance to reheating \cite{gorbunov}, we leave such cosmological interpretations to future work. This would be interesting in the case that gravitational wave experiments like LISA find evidence for a first order cosmological phase transition \cite{Boileau:2022ter}.

An important simplification that we made was to consider vanishing chemical potential. This was in part to simplify the presentation as the $U(1)$ phase plays little role in the overall dynamics of the system. However, looking ahead to the full $O(4)$ case necessary for the description of the chiral phase transition \cite{Grossi:2020ezz,Grossi:2021gqi,Florio:2021jlx,Soloviev:2022mte,Florio:2023kmy}, the non-Abelian phase could have non-trivial dynamics. Since the phase would have the interpretation of the different species of pions, this could be an important input in the predicted abundances of pion production \cite{Grossi:2021gqi}.

Looking further ahead, it would be interesting to explore superfluids in a UV-complete theory, outside of the hydrodynamic approximation. This can be implemented by including a sector with spontaneous symmetry breaking in a strongly coupled system, such as the well-known holographic Bjorken flow \cite{Janik:2005zt,Kinoshita:2008dq} and the recently developed holographic Gubser flow
\cite{Banerjee:2023djb,Mitra:2024zfy}. In the same vein, our present work has hydrodynamic transport coefficients as constants, whereas  finite coupling corrections to transport coefficients in holographic models are known \cite{Grozdanov:2016zjj}. Another option would be study weakly coupled kinetic theory in the relaxation time approximation 
\cite{Romatschke:2015gic, Kurkela:2017xis,Bajec:2024jez}, where hydrodynamic attractors have been previously studied \cite{Romatschke:2017acs,Kurkela:2019set}.

\begin{acknowledgments}
We would like to thank Matej Bajec, Eduardo Grossi, Sa{\v s}o Grozdanov, Micha\l~Heller, Ayan Mukhopadhyay, Enrico Pajer and Alexandre Serantes for helpful discussions.
G.K.B. acknowledges support for master's studies from the Republic of Slovenia (MVZI) and the European Union - NextGenerationEU (SiQUID-101091560) and the Ad Futura scholarship, Public Call no.~268 funded by the Public Scholarship, Development, Disability and Maintenance Fund of the Republic of Slovenia (JSRIPS). T.M.~has been supported by an appointment to the JRG Program at the APCTP through the Science and Technology Promotion Fund and Lottery Fund of the Korean Government, by the Korean Local Governments – Gyeongsangbuk-do Province and Pohang City – and by the National Research Foundation of Korea (NRF) funded by the Korean government (MSIT) (grant number 2021R1A2C1010834).
A.S. was supported by funding from Horizon Europe research and innovation programme under the Marie Skłodowska-Curie grant agreement No. 101103006 and the project N1-0245 of Slovenian Research Agency (ARIS).
\end{acknowledgments}

\bibliography{super}

\end{document}